\documentclass[a4paper, 11pt]{article}
\usepackage[T1]{fontenc}
\usepackage[utf8]{inputenc}
\usepackage[english]{babel}
\usepackage{amsmath}
\usepackage{amssymb}
\usepackage{braket}
\usepackage{mathtools}
\usepackage{dsfont}
\usepackage{subcaption}
\usepackage{array}
\usepackage{booktabs}
\usepackage{yfonts}
\usepackage{setspace}
\usepackage{verbatim}

\usepackage[usenames,dvipsnames]{color}

\numberwithin{equation}{section}

\def\be{\begin{equation}}
\def\ee{\end{equation}}

\newcommand{\diff}{\mathrm{d}}

 \newcommand{\zp}{{f}_{p}}
\newcommand{\zm}{{f}_{w}}
\newcommand{\zz}{{g}}
\definecolor{applegreen}{rgb}{0.55, 0.71, 0.0}

\usepackage{jheppub}                   

\makeatletter
\gdef\@fpheader{\ }                    
\makeatother

\title{Revisiting $\alpha'$ corrections to heterotic two-charge black holes}

\author[a]
{Stefano Massai,}
\author[b]
{Alejandro Ruip\'erez,}
\author[c]
{Matteo Zatti,}
\emailAdd{stefano.massai@pd.infn.it, alejandro.ruiperez@roma2.infn.it, matteo.zatti01@uam.es}

\affiliation[a]{Dipartimento di Fisica e Astronomia ``Galileo Galilei'', Universit\`a di Padova \& INFN Padova, Via Marzolo 8, 35131 Padova, Italy}
\affiliation[b]{Dipartimento di Fisica, Universit\`a di Roma ``Tor Vergata'' \& INFN  Roma 2,\\ Via della Ricerca Scientifica 1, 00133, Roma, Italy}
\affiliation[c]{Instituto de F\'isica Te\'orica UAM/CSIC,\\
 C/ Nicol\'as Cabrera, 13-15, C. U. Cantoblanco, E-28049 Madrid, Spain}

\abstract{We find solutions of the heterotic string effective action describing the first-order $\alpha'$ corrections to two-charge black holes at finite temperature. Making explicit use of these solutions, we compute the corrections to the thermodynamic quantities: temperature, chemical potentials, mass, charges and entropy. We check that the first law of black hole mechanics is satisfied and that the thermodynamics agrees with the one extracted from the Euclidean on-shell action. Finally, we show that our results are in agreement with the corrections for the thermodynamics recently predicted by Chen, Maldacena and Witten.}

\begin{document}

\begin{flushright}
{\tt \small{IFT-UAM/CSIC-23-143}} \\
\end{flushright}

\allowdisplaybreaks
\maketitle

\newpage

\section{Introduction}

The study of two-charge black holes has attracted much attention since the first investigations of black holes in string theory. This is mainly due to the fact that they are supposed to describe perhaps the simplest configuration in string theory which has a non-vanishing degeneracy of BPS states. This microscopic system consists of a fundamental heterotic string with winding $Q_{w}$ and momentum $Q_p$ charges along a compact direction ${\mathbb S}^{1}_{y}$. The degeneracy of BPS states of this system was computed by Dabholkar and Harvey \cite{Dabholkar:1989jt, Dabholkar:1990yf}, and it is given by 
\begin{equation}
S_{\rm{micro}}(Q_p, Q_w)=\log d\left(Q_p, Q_w\right)=4\pi \sqrt{Q_p Q_w}\,.
\end{equation}
 Being a BPS degeneracy, it must be protected when extrapolating it to the finite string-coupling regime where an effective black hole description is expected to exist (a priori). In other words, it should be possible to match this BPS degeneracy with the Bekenstein-Hawking entropy of the corresponding black hole. However, when trying to do so one finds a puzzle: even though there is a supergravity solution with the same charges and preserving the same supersymmetries as the Dabholkar-Harvey states \cite{Sen:1994eb, Cvetic:1995uj, Dabholkar:1995nc, Callan:1995hn}, it describes a singular black hole with vanishing horizon area. Hence, the naive macroscopic entropy that can be associated to the two-charge system vanishes. 

In order to explain this mismatch, Sen proposed \cite{Sen:1995in} that two-charge black holes have a small horizon of string size, and therefore it cannot be resolved by supergravity unless the latter is supplemented with higher-derivative terms capturing stringy $\alpha'$ corrections.\footnote{For this reason, these black holes are often referred to as \emph{small} black holes.} Almost ten years after this proposal, it was claimed in \cite{Dabholkar:2004yr, Dabholkar:2004dq} that four-derivative corrections in the context of type IIA on ${\mathbb K}_{3}\times {\mathbb T}^2$ (which is dual to heterotic on ${\mathbb T}^6$) stretch the horizon of two-charge black holes (hiding the singularity behind) and, what is even more remarkable, also give the precise contribution to the black hole entropy so that it reproduces the microstate counting of the two-charge system. 

These results, however, have been recently questioned in a series of papers \cite{Cano:2018hut, Ruiperez:2020qda, Cano:2021dyy} in which, working directly within the heterotic theory, it has been shown that $\alpha'$ corrections do not remove the singularity of BPS two-charge black holes. Furthermore, it has been argued that the configuration studied in \cite{Dabholkar:2004yr, Dabholkar:2004dq} should correspond to a regular four-dimensional black hole whose entropy accidentally matches the microscopic degeneracy of the two-charge system, but which carries different charges and preserves less supersymmetry. The fact that it preserves less supersymmetry is indeed a smoking gun of the presence of additional sources (NS5 branes and Kaluza-Klein monopoles), which would be the ultimate reason explaining why this four-dimensional black hole has a regular horizon. 

The fact that the two-charge system does not seem to admit a black hole description in the BPS limit is something which appears rather natural from the point of view of the correspondence between black holes and fundamental strings \cite{Susskind:1993ws, Horowitz:1996nw, Horowitz:1997jc, Damour:1999aw} (see also \cite{Chen:2021emg, Chen:2021dsw, Brustein:2021cza, Matsuo:2022kvx, Balthazar:2022hno, Ceplak:2023afb} for recent discussions). According to this proposal, black holes should turn into highly-excited strings when their sizes are of the order of the string scale. This has been recently discussed by Chen, Maldacena and Witten \cite{Chen:2021dsw} precisely in the context of the two-charge system. Let us consider a two-charge black hole at finite temperature. This can be described in supergravity in terms of a solution with a large (macroscopic) horizon. However, if the black hole starts losing its mass it will reach the string size before reaching extremality, which would imply that the right description of the system near extremality should be a sort of self-gravitating string solution \cite{Horowitz:1997jc, Damour:1999aw} rather than a solution with a horizon \cite{Chen:2021dsw}.\footnote{See also \cite{Mathur:2018tib} for a complementary point of view on this.}

In this paper we will mainly focus on two-charge black holes at finite temperature.
More concretely, we consider two-charge black holes in heterotic string theory and we study how the first-order $\alpha'$ corrections modify the solutions and their thermodynamic properties. The corrections to the thermodynamics have been recently studied in \cite{Chen:2021dsw}, exploiting the fact that the two-charge solutions can be obtained by perfoming suitable $O(2, 2)$ transformations to the Schwarzschild-Tangherlini solution, whose $\alpha'$ corrections had been already studied in \cite{Callan:1988hs}. In principle this method can be used not only to obtain the corrections to the thermodynamics but also the corrected solutions themselves, which were not provided in \cite{Chen:2021dsw}. This is just technically more involved, as one would have to take into account that the  $O(2, 2)$ transformations receive $\alpha'$ corrections \cite{Bergshoeff:1995cg, Kaloper:1997ux, Bedoya:2014pma, Ortin:2020xdm, Eloy:2020dko}.\footnote{As explained in \cite{Chen:2021dsw}, one can ignore the explicit corrections to the $O(2, 2)$ transformations if the goal is just to obtain the corrected thermodynamics.} This was precisely the strategy followed in \cite{Giveon:2009da}. However, as pointed out in \cite{Chen:2021dsw}, the corrected thermodynamics obtained in these two references do not agree between each other.  Our main motivation here is to perform an independent “first principle” computation of the corrected solutions and their thermodynamics; we are going to find the corrected solutions explicitly by solving the $\alpha'$-corrected equations of motion and then compute the thermodynamic quantities with standard methods.

Anticipating our results, we are going to show that the $\alpha'$ corrections to the thermodynamics that we compute fully agree with those of \cite{Chen:2021dsw}. This is a strong consistency check of both approaches, as well as of the methods employed and of the results obtained in previous related works by two of the authors and collaborators, see e.g.~\cite{Cano:2018hut, Ruiperez:2020qda, Cano:2021dyy, Cano:2018qev, Chimento:2018kop, Cano:2018brq, RuiperezVicente:2020qfw, Elgood:2020nls, Elgood:2020xwu, Cano:2021nzo, Ortin:2021win, Cano:2022tmn, Zatti:2023oiq} and references therein. In particular, we want to emphasize that the (singular) solutions found in \cite{Cano:2018hut, Ruiperez:2020qda, Cano:2021dyy} are properly recovered from the non-extremal ones we have found in this paper after suitably taking the extremal limit. This is discussed in subsection~\ref{sec:BPS_limit} and  further confirms the conclusions of \cite{Cano:2018hut, Ruiperez:2020qda, Cano:2021dyy}, yet from a different perspective.

The organization of the rest of the paper is the following. In section~\ref{sec:solutions} we review the two-derivative solutions describing heterotic two-charge black holes in arbitrary dimension ($4\le d\le 9$) and then provide the details about the $\alpha'$-corrected solutions, focusing on the four- and five-dimensional cases. In section~\ref{sec:BH_thermodynamics} we compute the thermodynamic quantities of the solutions and express them using two-different parametrizations: fixing the value of the mass and charges (micro-canonical ensemble) and fixing the inverse temperature and the chemical potentials (grand-canonical ensemble). We show that the results we get are consistent with the first law of black hole mechanics. Then in section~\ref{sec:chargesfromI} we corroborate the results of section~\ref{sec:BH_thermodynamics} by employing an alternative method to compute the corrected thermodynamics, namely from the Euclidean on-shell action. Finally, in section~\ref{sec:2QsBHfromST} we compare our results with those of \cite{Chen:2021dsw} finding that they are in perfect agreement. The appendices contain additional information on the effective action and equations of motion in app.~\ref{app:eff_action+EOMs}, on the procedure followed to find the corrected solutions in app.~\ref{app:correctedsol}, and on the dimensional reduction of the configurations on ${\mathbb S}^1_y$ in app.~\ref{app:dimensional_red}.

\textbf{Note on conventions.} We adopt the conventions of \cite{Ortin:2015hya}. In particular, we use the mostly minus signature for the metric $(+ - \dots -)$ and the conventions for the Riemann tensor are such that
\begin{equation}
[\nabla_{\mu},\nabla_{\nu}]\xi^{\sigma}=R_{\mu\nu\rho}{}^{\sigma}\xi^{\rho}\,,
\end{equation}
The conventions for the remaining fields are specified in appendix~\ref{app:eff_action+EOMs}. Hats $\hat {\cdot}$ shall be used to denote both ten- and $(d+1)$-dimensional fields, as they are identified with one another. Instead, the $d$-dimensional fields which are obtained upon compactification on ${\mathbb S}^{1}_{y}$ will not carry hats.

\section{$\alpha'$ corrections to heterotic two-charge black holes}\label{sec:solutions}

\subsection{Two-charge black holes at leading order in $\alpha'$}

Let us begin reviewing the two-derivative solution describing non-extremal two-charge black holes in $d$ dimensions \cite{Horowitz:1996nw}. Given that in subsection~\ref{sec:alpha_corrected_BHs} we will solve the corrected ten-dimensional equations of motion, here we directly present the solution in its ten-dimensional form. However, since the solutions have a ${\mathbb T}^{(9-d)}$ torus playing a trivial role, we feel free to ignore these torus directions from now on.\footnote{Taking them into account just amounts to add the flat metric on the torus $-d{\vec z}^2_{(9-d)}$ to the $(d+1)$-dimensional metric \eqref{eq:metric}.} Doing so, the resulting $(d+1)$-dimensional solution is given by 
\begin{eqnarray}
\label{eq:metric}
{\diff}{\hat s}^2&=&\frac{f}{\zp \zm}\,{\diff}t^2-f^{-1}{\diff}\rho^2-\rho^2 {\diff}\Omega^2_{(d-2)}-k^2_{\infty}\frac{\zp}{\zm}\left({\diff}y+\beta_pk^{-1}_{\infty}\left(\zp^{-1}-1\right){\diff}t\right)^2\,,\\
{\hat B}&=&\beta_w k_{\infty}\left(\zm^{-1}-1\right) {\diff}t \wedge {\diff}y\,, \\
e^{2\hat \phi}&=&e^{2\hat \phi_{\infty}}\zm^{-1}\,,
\end{eqnarray}
where $\diff{\hat s}$ represents the line element  in the string frame and
\begin{equation}
\zp=1+\frac{q_{p}}{\rho^{d-3}}\, , \hspace{1cm}\zm=1+\frac{q_{w}}{\rho^{d-3}}\, , \hspace{1cm} f=1-\frac{\rho_s^{d-3}}{\rho^{d-3}}\, .
\end{equation} 
The parameters $q_p$, $q_w$ and $\rho_s$ are related to the charges and mass of the solutions. Together with the moduli ${\hat \phi}_{\infty}$ and $k_{\infty}$  (representing the asymptotic values of the dilaton and the Kaluza-Klein scalar), they constitute the set of independent parameters of the solutions since $\beta_{p}$ and $\beta_{w}$ are subject to the following constraints,
\begin{equation}
\rho_s^{(d-3)}=q_{p}\left(\beta^2_{p}-1\right)=q_{w}\left(\beta^2_{w}-1\right)\, ,
\end{equation}
implying that
\begin{equation}
\beta_{i}=\epsilon_{i}\sqrt{1+ \frac{\rho_s^{d-3}}{q_{i}}}\, , \hspace{1cm} i=\left\{p, w\right\}\, .
\end{equation}
where $\epsilon^2_{i}=1$.  These correspond to the signs of the winding and momentum charges, respectively. In the BPS limit ($\rho_s\to 0$), the solution with $\epsilon_{w}=\epsilon_{p}$ is supersymmetric, while the one with $\epsilon_{w}=-\epsilon_{p}$ does not preserve any supersymmetry. The analysis of the Killing spinor equations for these configurations can be found for instance in \cite{Ruiperez:2020qda, Cano:2021dyy, Cano:2021nzo}.

\subsection{$\alpha'$-corrected solutions}
\label{sec:alpha_corrected_BHs}

Our aim now is to compute the first-order $\alpha'$ corrections to these two-charge black holes. As usual, we treat the $\alpha'$ corrections in a perturbative fashion and ignore ${\cal O}(\alpha'^2)$ terms. The first-order $\alpha'$ corrections in the effective action of the heterotic superstring were studied in \cite{Gross:1986mw, Metsaev:1987zx, Bergshoeff:1989de}. While different approaches were used, it was later shown in \cite{Chemissany:2007he} that the resulting effective actions are equivalent up to field redefinitions. Here we choose to work in the Bergshoeff-de Roo scheme, \cite{Bergshoeff:1989de}. In order to establish our conventions, we review the effective action and equations of motion in appendix~\ref{app:eff_action+EOMs}.
 
Before entering into the details of the corrected solutions, let us briefly explain the general strategy we have followed in order to find the corrected solutions. The interested reader is referred to appendix~\ref{app:correctedsol} or to \cite{Cano:2022tmn} for more details.  It turns out that an educated ansatz to solve the corrected equations of motion is the following,
\begin{eqnarray}
\label{eq:ansatz_metric}
{\diff}{\hat s}^2&=&\frac{f}{\zp  {\tilde f}_{w}}{\diff}t^2-\zz\left(f^{-1}{\diff}\rho^2+\rho^2 {\diff}\Omega^2_{(d-2)}\right)-k^2_{\infty}\frac{\zp}{ {\tilde f}_{w}}\left[{\diff}y+\beta_{p}k^{-1}_{\infty}\left(\zp^{-1}-1\right){\diff}t\right]^2\,,\\[1mm]
\label{eq:ansatz_B}
{\hat B}&=&\beta_w k_{\infty}\left({f}_{w}^{-1}-1\right) {\diff}t \wedge {\diff}y\,,
\end{eqnarray}
where the functions  $f, f_p, f_w, {\tilde f}_w, g$ and the dilaton $\hat \phi$ are assumed to depend only on the radial coordinate $\rho$. For consistency with the perturbative approach, they must be of the form
\begin{equation}
\begin{aligned}
{f}_{p}=\,&1+\frac{q_p}{\rho^{d-3}}+\alpha' {\delta f_p}\, , \hspace{5mm} {\tilde f}_{w}=\,1+\frac{q_w}{\rho^{d-3}}+\alpha' {\delta {\tilde f}_{w}}\, ,\hspace{5mm}\zz=\,1+\alpha' {\delta \zz}\, ,\\[1mm]
{f}=&\,1-\frac{\rho_s^{d-3}}{\rho^{d-3}}+\alpha' {\delta f}\, ,\hspace{5mm} {f}_{w}=\,1+\frac{q_w}{\rho^{d-3}}+\alpha' {\delta {f}_{w}}\, .
\end{aligned}
\end{equation}
After linearization in $\alpha'$, the equations of motion boil down to a linear system of inhomogeneous second-order ODEs for the unknown functions $\delta f_p, \delta {\tilde f}_w, \delta g, \delta f, \delta f_w$ and $\hat \phi$. The strategy we are going to follow to solve them is the same as in \cite{Cano:2022tmn}, which consists of performing an asymptotic expansion (large $\rho$) of the unknown functions and solve the equations of motion order by order. Following this procedure, we can determine all the coefficients of the asymptotic expansion except for a few of them which remain free, the integration constants.  Once the form of the asymptotic solution has been found, we resum the asymptotic series with the help of \texttt{Mathematica}. The final step is to fix the integration constants by imposing regularity at the horizon and suitable boundary conditions. Our choice here will be such that we keep the asymptotic charges and the mass fixed: \textit{i.e.}, we are going to give the form of the corrected solution in the micro-canonical ensemble.

In what follows we give the corrected solutions in $d=5$ and $d=4$, as well as its dimensional reduction on ${\mathbb S}^{1}_{y}$. Finally, we study their BPS limits and check that they agree with the corrected solutions found in \cite{Cano:2018hut, Ruiperez:2020qda, Cano:2021dyy}.

\subsubsection{Five-dimensional black holes}
Let us first consider the $d=5$ case.  The expression of the dilaton is found by solving the equation of motion of the Kalb-Ramond 2-form $\hat B$. It is given by
\begin{equation}
\hat \phi=a_{\phi}+\frac{1}{2}\log \frac{-\rho^3 \zz {\tilde f}_w {f}'_{w}}{2q_{w}{f}^2_{w}}\, ,
\end{equation}
where $a_{\phi}$ is an integration constant to be fixed by imposing the asymptotic value of the string coupling is not renormalized, namely
\begin{equation}
\lim_{\rho\to\infty}{\hat \phi}={\hat \phi}_{\infty}\, .
\end{equation}
After fixing the integration constants in the way we have explained, we find the following solution:
\begin{eqnarray}
\delta \zp&=&-\frac{q_p \rho _s^2 \left(1+\frac{\beta_w}{\beta_p}\right) \log \left(1+\frac{q_w}{\rho ^2}\right)}{2 q_w^2 \rho ^2  }\nonumber\\
&+&\frac{1}{32 q_w \left(\rho ^2+q_w\right) \rho ^6 }\left\{16 q_p \rho_s^2 \rho ^4-q_w q_p \rho ^2 \left(9 \rho _s^2+32 q_w\right)+7 q_w^2 q_p \rho_s^2 \right.\nonumber\\
&+&\left.\frac{\beta_w}{\beta_p}\left[16 q_p \rho_s^2 \rho ^4+8 q_w \rho ^2 \left(q_p \rho_s^2-2 q_w \left(\rho_s^2+2 q_p\right)\right)+8 q_w^2 q_p \rho_s^2\right]\right\}\,,\\[4mm]
\delta {\tilde f}_w&=&-\frac{\rho_s^2 \left(1+\frac{\beta_w}{\beta_p}\right) \log \left(1+\frac{q_w}{\rho ^2}\right)}{2 q_w\rho ^2 }+\frac{\beta_w\rho_s^2 \left(q_w+2 \rho^2\right) }{4\beta_p\rho^6}+\frac{7 q_w \rho_s^2}{32\rho^6}+\frac{\rho_s^2}{2 \rho^4}\nonumber\\
&+& \frac{9\rho_s^2 (q_p-q_w)}{4 \rho^2 \left(4 \rho_s^2 (q_w+q_p)+4 q_wq_p+3 \rho_s^4\right)}\,,\\[4mm]
\delta \zz&=&\frac{\rho_s^2 \left(1+\frac{\beta_w}{\beta_p}\right) \log \left(1+\frac{q_w}{\rho ^2}\right)}{2 q_w^2}+\frac{\beta_w\rho _s^2 \left(q_w-2 \rho ^2\right)}{4\beta_p q_w \rho ^4 }-\frac{7\rho_s^2}{32 \rho^4}\nonumber\\
&+&\frac{1}{8 \rho^2} \left(-\frac{4 \rho_s^2}{q_w}+\frac{18 q_p \left(\rho_s^2+2 q_w\right)}{4 \left(q_w+q_p\right) \rho_s^2+3 \rho_s^4+4 q_w q_p}-9\right)\,,\\[4mm]
\delta f&=&\frac{\rho_s^4 \left(1+\frac{\beta_w}{\beta_p}\right) \log \left(1+\frac{q_w}{\rho ^2}\right)}{2 q_w^2\rho ^2 }-\frac{\beta_w \rho_s^2 \left(2 \rho_s^2 \rho ^4+q_w \rho_s^2 \rho ^2+q_w^2 \left(3 \rho_s^2-4 \rho ^2\right)\right)}{4q_w \beta_p\rho^6\left(\rho ^2+q_w\right) }\nonumber\\
&+&\frac{\rho_s^2q_w }{\left(\rho ^2+q_w\right)\rho^4}-\frac{3 \left(\rho ^2+q_p\right)}{4 \rho^4}+\frac{\rho_s^4 \left(-16 \rho ^4+9 q_w \rho ^2-7 q_w^2\right)}{32q_w \rho^6\left(\rho ^2+q_w\right)}\nonumber\\
&+&\frac{ \left(\rho ^2+q_p\right) \left(2 q_w q_p-\left(q_w-2 q_p\right) \rho_s^2\right)}{2\rho^4\left(4 \left(q_w+q_p\right) \rho_s^2+3 \rho_s^4+4 q_w q_p\right)}\, ,\\[4mm]
\delta {f}_{w}&=&-\frac{\rho_s^2 \left(1+\frac{\beta_w}{\beta_p}\right) \log \left(1+\frac{q_w}{\rho ^2}\right)}{2 q_w\rho ^2 }-\frac{\beta_w\rho_s^2 \left(q_w-2 \rho ^2\right)}{4\beta_p  \rho ^6} -\frac{q_w\rho_s^2}{32 \rho^6}\nonumber\\
&+&\frac{\rho_s^2 \left(4+\frac{18 q_w \left(q_w-q_p\right)}{4 \left(q_w+q_p\right) \rho_s^2+3 \rho_s^4+4 q_w q_p} \right)}{8 \rho^4}\, ,
\end{eqnarray}
and $a_{\phi}={\hat \phi}_{\infty}$.

\subsubsection{Four-dimensional black holes}

We proceed as in the five-dimensional case. Solving the ten-dimensional Kalb-Ramond equation of motion yields the expression for the dilaton,
\begin{equation}
	\hat{\phi} = \hat{\phi}_\infty + \frac{1}{2}\log \frac{-\rho^2 {\zz}^{1/2} {\tilde f}_w {f}'_{w}}{q_{w}{f}^2_{w}}\,,
\end{equation}
where we have already imposed that the asymptotic value of the string coupling is not renormalized. Solving the remaining equations of motion as explained before we obtain: 
\begin{eqnarray}
	\delta \zp &=& \frac{\beta_p\beta_wq_p \left[q_p \left(4 q_w^3 (\rho_s-3 \rho)+\rho_s q_w^2 \rho-3 \rho_s q_w \rho^2-6 \rho_s \rho^3\right)-6 \rho_s q_w^3 \rho\right]}{48 q_w^2 \rho^4 (q_p+\rho_s) (q_w+\rho)} \nonumber\\
	&-& \frac{q_p \left[q_w^3 (39 \rho-10 \rho_s)+q_w^2 \rho (17 \rho_s+3 \rho)+9 \rho_s q_w \rho^2+18 \rho_s \rho^3\right]}{144 q_w^2 \rho^4 (q_w+\rho)}\nonumber\\
	&+& \log \left(1+\frac{q_w}{\rho}\right) \left(\frac{\beta_p\beta_wq_p^2 \rho_s}{8 q_w^3 \rho (q_p+\rho_s)}+\frac{q_p \rho_s}{8 q_w^3 \rho}\right) \, ,\\[4mm]
	\delta {\tilde f}_w &=& \frac{\beta_p\beta_wq_p \rho_s \left(4 q_w^2+3 q_w \rho-6 \rho^2\right)}{48 q_w \rho^4 (q_p+\rho_s)} + \frac{q_w^2 (10 \rho_s-3 \rho)+9 \rho_s q_w \rho-18 \rho_s \rho^2}{144 q_w \rho^4} \nonumber\\
	&+& \frac{q_p-q_w}{\rho \left(4 q_p q_w +3 q_p \rho_s +3 \rho_s q_w +2 \rho_s^2 \right)} \nonumber\\
	&+& \log \left(1+\frac{q_w}{\rho}\right)\left(\frac{\beta_p\beta_wq_p \rho_s}{8 q_w^2 \rho (q_p+\rho_s)}+\frac{\rho_s}{8 q_w^2 \rho}\right) \,,\\[4mm]
	\delta \zz &=& \frac{\beta_p\beta_wq_p \rho_s \left(2 q_w^2-3 q_w \rho+6 \rho^2\right)}{24 q_w^2 \rho^3 (q_p+\rho_s)} -\frac{\rho_s+3 q_w}{8 q_w \rho^2}-\frac{5 \rho_s}{36 \rho^3} \nonumber\\
	&-& \frac{(\rho_s+2 q_w) \left(-3 q_p \rho_s+2 q_p q_w-2 \rho_s^2+6 q_w^2+\rho_s q_w\right)}{4 q_w^2 \rho \left(3 q_p \rho_s+4 q_p q_w+2 \rho_s^2+3 \rho_s q_w\right)}  \nonumber\\
	&-& \log \left(1+\frac{q_w}{\rho}\right) \left(\frac{\beta_p\beta_wq_p \rho_s}{4 q_w^3 (q_p+\rho_s)}+\frac{\rho_s}{4 q_w^3}\right) \,,\\[4mm]
	\delta f &=& \frac{\beta_p\beta_wq_p \rho_s \left(2 q_w^3 (6 \rho-5 \rho_s)-\rho_s q_w^2 \rho+3 \rho_s q_w \rho^2+6 \rho_s \rho^3\right)}{48 q_w^2 \rho^4 (q_p+\rho_s) (q_w+\rho)} \nonumber\\
	&-& \frac{40 q_p \rho_s^2 q_w^4+30 \rho_s^3 q_w^3 (q_p+q_w)+20 \rho_s^4 q_w^3+\rho^4 \left(288 q_w^3+144 \rho_s q_w^2\right)}{144 q_w^2 \rho^4 (q_w+\rho) \left[3 \rho_s (q_p+q_w)+4 q_p q_w+2 \rho_s^2\right]} \nonumber\\
	&+& \frac{\rho^3 \left(-288 q_w^3 (q_p+q_w)-144 \rho_s q_w^2 (q_p+q_w)+54 \rho_s^3 (q_p+q_w)+72 q_p \rho_s^2 q_w+36 \rho_s^4\right)}{144 q_w^2 \rho^4 (q_w+\rho) \left[3 \rho_s (q_p+q_w)+4 q_p q_w+2 \rho_s^2\right]} \nonumber\\
	&+& \frac{\rho^2 \left[9 \rho_s^2 q_w^2 (5 q_p+q_w)+3 \rho_s^3 q_w (9 q_p+11 q_w)+18 \rho_s^4 q_w -288 q_p q_w^4-132 q_p \rho_s q_w^3 \right]}{144 q_w^2 \rho^4 (q_w+\rho) \left[3 \rho_s (q_p+q_w)+4 q_p q_w+2 \rho_s^2\right]} \nonumber\\
	&+&  \frac{\rho \left[156 q_p \rho_s q_w^4+\rho_s^2 q_w^3 (185 q_p+117 q_w)+3 \rho_s^3 q_w^2 (17 q_p+43 q_w)+34 \rho_s^4 q_w^2\right]}{144 q_w^2 \rho^4 (q_w+\rho) \left[3 \rho_s (q_p+q_w)+4 q_p q_w+2 \rho_s^2\right]} \nonumber\\
	&-& \log \left(1+\frac{q_w}{\rho}\right) \left(\frac{\beta_p\beta_wq_p \rho_s^2}{8 q_w^3 \rho (q_p+\rho_s)}+\frac{\rho_s^2}{8 q_w^3 \rho}\right) \, ,\\[4mm]
	\delta {f}_{w} &=& -\frac{\beta_p\beta_wq_p \rho_s \left(2 q_w^2-3 q_w \rho+6 \rho^2\right)}{48 q_w \rho^4 (q_p+\rho_s)}+\frac{q_w (q_w-q_p)}{\rho^2 \left(3 \rho_s q_p+ 3 \rho_s q_w +4 q_p q_w+2 \rho_s^2\right)} \nonumber\\
	&-& \frac{\rho_s \left(8 q_w^2-9 q_w \rho+18 \rho^2\right)}{144 q_w \rho^4}+\frac{q_w}{48 \rho^3}  \nonumber\\
	&+& \log \left(1+\frac{q_w}{\rho}\right) \left(\frac{\beta_p\beta_wq_p \rho_s}{8 q_w^2 \rho (q_p+\rho_s)}+\frac{\rho_s}{8 q_w^2 \rho}\right) \,. 
\end{eqnarray}

\subsection{Dimensional reduction on ${\mathbb S}^1_y$}

The dimensional reduction of this solution to five dimensions can be carried out using the formulae obtained in \cite{Elgood:2020xwu}, which are collected in appendix~\ref{app:dimensional_red}. Applying them to the configuration at hands, we get
\begin{eqnarray}
\diff s^2&=\,&\frac{f}{\zp  {\tilde f}_{w}}{\diff}t^2-\zz\left(f^{-1}{\diff}\rho^2+\rho^2 {\diff}\Omega^2_{(d-2)}\right)\, ,\\[1mm]
B&=\,&0\,,\\[1mm]
A&=\,&\beta_{p}\, k^{-1}_{\infty}\left(\zp^{-1}-1\right){\diff}t\,,\hspace{1cm} C=\,\beta_{w}\, k_{\infty}\left[\zm^{-1}\left(1+\alpha' \beta_w^{-1}\Delta_C \right)-1\right]{\diff}t\,,\\[1mm]
e^{2\phi}&=&e^{2\hat\phi}\,k^{-1}_{\infty} \left(\frac{{\tilde f}_w}{f_p}\right)^{1/2}   \,, \hspace{1cm} k=k_{\infty}\left(\frac{f_p}{{\tilde f}_w}\right)^{1/2}\, ,
\end{eqnarray}
where $\diff s$ represents the line element in the string frame and 
\begin{equation}\label{eq:DeltaC}
\Delta_{C}=\frac{2\left(\beta_p+\beta_w\right)f f'_p f'_w-f'\left(\beta_w f_p f'_w+\beta_p f'_p f_w\right) }{8f_p f_w} .
\end{equation}

\subsection{Extremal limit}\label{sec:BPS_limit}

The extremal limit is implemented by setting $\rho_s\to 0$ while keeping the charge parameters $q_p$ and $q_w$ fixed. The $\alpha'$ corrections in this limit have been already studied in the recent literature \cite{Cano:2018hut, Ruiperez:2020qda, Cano:2021dyy}. The corrected solution in arbitrary dimension is given by \cite{Ruiperez:2020qda, Cano:2021dyy}:\footnote{Here we are focusing on the supersymmetric case $\epsilon_{p}=\epsilon_{w}$, which was the case analyzed in \cite{Cano:2018hut, Ruiperez:2020qda, Cano:2021dyy}. Surprisingly, in the non-supersymmetric case $\epsilon_{p}=-\epsilon_{w}$ the corrections simply vanish as the first-order correction in \eqref{eq:fp} is multiplied by $1+\epsilon_p \epsilon_w$.}
\begin{eqnarray}
f&=&g=1\,,\\[1mm]
\label{eq:fp}
f_{p}&=&1+\frac{q_p}{\rho^{d-3}}-\frac{(d-3)^2\alpha'}{2} \frac{q_p q_w}{\rho^{d-1}\left(\rho^{d-3}+q_w\right)}\,,\\[1mm]
{\tilde f}_{w}&=&f_{w}=1+\frac{q_w}{\rho^{d-3}}\,.
\end{eqnarray}
We have checked that this solution is precisely recovered from the non-extremal ones we have presented presented in subsection~\ref{sec:alpha_corrected_BHs} upon taking $\rho_s\to 0$.\footnote{While the limit is smooth in the five-dimensional case, in the four-dimensional one the limit must be taken before fixing the integration constants, as the expressions for the latter (which we have not provided explicitly) diverge when $\rho_s\to0$.} This is an interesting consistency check of our solutions.

\section{Black hole thermodynamics}\label{sec:BH_thermodynamics}

In this section we compute the thermodynamic quantities of the $\alpha'$-corrected solutions found in the previous section. Let us first explain how to compute them in general. Then we apply the corresponding formulae to the five- and four-dimensional solutions. 

\textbf{Temperature and chemical potentials.} The inverse temperature $\beta$ is obtained by demanding regularity in the Euclidean section \cite{Gibbons:1976ue}. A standard calculation shows that it is given by
\begin{equation}\label{eq:beta}
\beta= 4\pi \frac{\sqrt{g f_p {\tilde f}_{w}}}{f'}\Bigg|_{\rho=\rho_h}\, ,
\end{equation}
where $\rho_h$ is the position of the outer horizon. The latter corresponds to the (largest) root of the metric function $f$,
\begin{equation}
f(\rho_h)=0\, .
\end{equation}
As a consequence of our choice of boundary conditions the position of the horizon $\rho_h$ is shifted by the $\alpha'$ corrections.

The chemical potentials associated to the Kaluza-Klein and winding vectors are obtained as usual,
\begin{eqnarray}\label{eq:chem_pot}
\Phi_{p}=\xi^{\mu}A_{\mu}|_{\infty}-\xi^{\mu}A_{\mu}|_{\rho=\rho_h}\,, \hspace{1cm}\Phi_{w}=\xi^{\mu}C_{\mu}|_{\infty}-\xi^{\mu}C_{\mu}|_{\rho=\rho_h}\,,
\end{eqnarray}
where $\xi=\partial_t$ is the Killing vector that generates the horizon. As it turns out, the explicit correction obtained in \cite{Elgood:2020xwu} for the expression of the lower-dimensional winding vector $C_{\mu}$ (reported here in \eqref{eq:DeltaC}) plays a crucial role in order to obtain a result consistent with the first law of black hole mechanics. 

\textbf{Electric charges and mass.} The four-derivative corrections modify the two-derivative Maxwell equations of the vector fields, giving rise to several notions of charge \cite{Page:1983mke, Marolf:2000cb}. However, we expect on general grounds that all of them should coincide when computed asymptotically, provided the field strengths (and other quantities involved) decay fast enough. Given this, we compute the so-called Maxwell charges \cite{Marolf:2000cb}:\footnote{Our conventions for the Hodge star operator are such that $\star \left(dx^{\mu_1}\wedge \dots \wedge dx^{\mu_n}\right)=\frac{1}{(d-n)!}\epsilon_{\nu_1 \dots \nu_{d-n}}{}^{\mu_1 \dots \mu_n}\, dx^{\nu_1}\wedge \dots \wedge dx^{\nu_{d-n}}$ and $\epsilon_{01\dots d-1}=+\sqrt{|g|}$.}
\begin{equation}
Q_{p}= \frac{-1}{16\pi G_N}\int_{{\mathbb S}^{(d-2)}_{\infty}}e^{-2\left(\phi-\phi_{\infty}\right)}k_{(1)}^2 \star F \,, \hspace{5mm}Q_{w}=\frac{-1}{16\pi G_N}\int_{{\mathbb S}^{(d-2)}_{\infty}}e^{-2\left(\phi-\phi_{\infty}\right)}k^{-2} \star G \, ,
\end{equation}
where $F=\diff A$, $G=\diff C$, $k_{(1)}$ is the scalar combination given in \eqref{eq:scalarcombination} and $G_N$ is the $d$-dimensional Newton constant,
\begin{equation}
G_N=\frac{\hat G_N}{2\pi R_{y}}=\frac{\hat G_N}{2\pi k_{\infty}\ell_s}\, ,
\end{equation}
being $\hat G_N$ the $(d+1)$-dimensional one. 

The expression for the mass $E$ can be obtained by applying the ADM formula. In practice, we can just identify $E$ by looking at the asymptotic behavior of the $tt$ component of the metric in the modified Einstein frame \cite{Maldacena:1996ky},
\begin{equation}
g_{E}{}_{tt}=e^{-\frac{4}{d-2}(\phi-\phi_{\infty})} g_{tt}\approx 1- \frac{16\pi G_N E}{(d-2)\omega_{(d-2)}\rho^{d-3}}+\dots \,,
\end{equation}
where $\omega_{(d-2)}$ is the volume of the unit ${\mathbb S}^{(d-2)}$ sphere.

\textbf{Black hole entropy.}  In higher-derivative theories the entropy follows from Wald's formula \cite{Wald:1993nt, Iyer:1994ys}. However, one of the key assumptions in its derivation does not hold in presence of gravitational Chern-Simons terms, such as the ones present in the heterotic theory.\footnote{Namely, that the transformation of the $d$-form Lagrangian $\mathbf L$ under diffemorphisms is $\delta_{\xi}{\mathbf L}={\cal L}_{\xi}{\mathbf L}$, being ${\cal L}_{\xi}$ the Lie derivative with respect to $\xi$. This property is however satisfied by mixed Chern-Simons terms of the form $A_{(p)}\wedge {\rm{tr}}\left(R\wedge R\right)$, where $A_{(p)}$ is some $p$-form potential. In this case one can show \cite{Cassani:2023vsa} that Wald's formula is still applicable if suitable regularity conditions are imposed on $A_{(p)}$.} As a consequence, different strategies have been proposed in the literature in order to circumvent this issue (see e.g.~\cite{Sahoo:2006vz, Sahoo:2006pm, Faedo:2019xii, Elgood:2020xwu, Ortin:2020xdm, Ma:2022nwq} for a limited list of references), which mainly involve a convenient rewriting of the action. Nevertheless, it is also possible to extend Wald's formalism to properly account for gravitational Chern-Simons terms. Doing so, general expressions for the black hole entropy were obtained in \cite{Tachikawa:2006sz} and more recently in \cite{Elgood:2020nls}. It is more convenient for us to  make use of the entropy formula given in \cite{Elgood:2020nls}, as it has been derived precisely in the context of the heterotic effective action. We report it here for completeness,
\begin{equation}
	\label{eq:Waldentropyformula}
	S
	=
	\frac{(-1)^{d+1} {\hat g}_{s}^{2}}{8\hat{G}_{N}}
	\int_{\mathcal{BH}}
	e^{-2\hat\phi}
	\left\{
	\left[
	\hat \star ({\hat e}^{\hat a}\wedge {\hat e}^{\hat b})
	+\frac{\alpha'}{2}\star \hat R_{(-)}{}^{\hat a \hat b}
	\right]{\hat n}_{\hat a\hat b}
	+(-1)^{d}\frac{\alpha'}{2}\Pi_{n}\wedge \hat\star {\hat H}
	\right\}\,,
\end{equation}
where $\mathcal{BH}$ stands for the bifurcation surface of the event horizon and $\hat R_{(-)}{}^{\hat a \hat b}$ is the curvature two-form defined in eq.~(\ref{eq:Rminus}). $\Pi_{n}$ is the vertical Lorentz momentum map associated to the binormal to the Killing horizon, $\hat n^{\hat a\hat b}$, and it is defined by the property
\begin{equation}
	d\Pi_{n}
	\stackrel{\mathcal{BH}}{=}
	{\hat R}_{(-)}{}^{\hat a\hat b}{\hat n}_{\hat a\hat b}\,.
\end{equation}
The formula for the entropy is gauge-invariant and frame independent. Performing a local Lorentz transformation we can always put the Vielbein components $\hat{e}_{\hat{\mu}}{}^{\hat{a}}$ in an upper triangular form. In such a frame and with our ansatz $\Pi_{n}$ has the explicit expression
\begin{equation}
	\Pi_{n}
	\stackrel{\mathcal{BH}}{=}
	{\hat \Omega}_{(-)}{}^{\hat a\hat b}{\hat n}_{\hat a\hat b}\,,
\end{equation}
and \eqref{eq:Waldentropyformula} can be easily evaluated.

Having explained how to compute the different thermodynamic quantities, we now apply the above formuale to find the corrected thermodynamics of two-charge black holes. In order to facilitate the comparison with the previous literature \cite{Giveon:2009da, Chen:2021dsw} (which is something that we will do later in section~\ref{sec:2QsBHfromST}), we introduce the notation which is used in the aforementioned references:
\begin{equation}
q_{i}=\rho_s^{d-3} \sinh^2\gamma_i\,, \hspace{1cm} i=\left\{p, w\right\}\, .
\end{equation}
In addition, we will write down the different expressions both in the micro- and grand-canonical ensembles. By definition, the first is the one in which the expressions for the mass $E$ and charges, $Q_{p}$ and $Q_{w}$, take the same form as in the two-derivative solution (the solution in section~\ref{sec:alpha_corrected_BHs} is given using this parametrization). In turn, what is fixed in the latter ensemble are the inverse temperature $\beta$ and the chemical potentials, $\Phi_{p}$ and $\Phi_{w}$.

\subsection{Thermodynamic quantities in the micro-canonical ensemble}

\textbf{Five-dimensional black holes.} By definition, the expressions for the charges are the same as in the two-derivative solution, namely
\begin{eqnarray}
Q_{p}&=&\frac{\epsilon_{p} k_{\infty} \pi}{8G_N} \rho_s^2 \sinh \left(2\gamma_p\right)\,, \\[1mm]
Q_{w}&=&\frac{\epsilon_{w} \pi}{8G_N k_{\infty}} \rho_s^2 \sinh \left(2\gamma_w\right)\,, \\[1mm]
E&=&\frac{\pi \rho _s^2}{8G_N} \left[1+\cosh \left(2 \gamma _p\right)+\cosh \left(2 \gamma
   _w\right)\right]\, .
\end{eqnarray}
Contrarily to the charges, the inverse temperature \eqref{eq:beta} and the chemical potentials \eqref{eq:chem_pot} receive $\alpha'$ corrections. Parametrizing them as follows,
\begin{eqnarray}
\beta&=&2\pi  \cosh \gamma_p \cosh\gamma_w \rho_s \left(1+\frac{\alpha' \Delta\beta}{\rho_s^2}\,\right)\,,\\[1mm]
\Phi_{p}&=&\frac{\epsilon_{p}\tanh\gamma_p}{k_{\infty}}\left(1+\frac{\alpha' \Delta \Phi_{p}}{\rho_s^2}\right)\,,\\[1mm]
 \Phi_{w}&=&\epsilon_{w}k_{\infty}\tanh\gamma_w\left(1+\frac{\alpha' \Delta \Phi_{w}}{\rho_s^2}\right)\,,
\end{eqnarray}
we get 
\begin{eqnarray}
\Delta \beta &=\,&-\frac{\epsilon_{p}\epsilon_w}{2}\tanh\gamma_p \tanh\gamma_w-\,\frac{9\left(4\sinh^2\gamma_p \sinh^2\gamma_w-1\right)}{8\left(4\cosh^2\gamma_p \cosh^2\gamma_w-1\right)}\,, \\[1mm]
\Delta\Phi_{p}&=\,&-\frac{\epsilon_{p}\epsilon_w \tanh \gamma_w}{\sinh \left(2\gamma_p\right)}-\frac{9\cosh\left(2\gamma_w\right)}{4\left(4\cosh^2\gamma_p \cosh^2\gamma_w-1\right)}\,, \\[1mm]
\Delta\Phi_{w}&=\,& -\frac{\epsilon_{p}\epsilon_w \tanh \gamma_p}{\sinh \left(2\gamma_w\right)}-\frac{9\cosh\left(2\gamma_p\right)}{4\left(4\cosh^2\gamma_p \cosh^2\gamma_w-1\right)}\,. 
\end{eqnarray}
Finally, the result that we obtain for the black hole entropy is
\begin{equation}
S=\frac{\pi^2 \rho_s^3\cosh\gamma_p \cosh\gamma_w }{2G_N}\left[1+\frac{\alpha'}{8\rho_s^2}\left(9+4\epsilon_{p}\epsilon_{w}\tanh \gamma_p \tanh \gamma_w\right)\right]\, .
\end{equation}
These expressions pass several consistency checks. First, one can verify that the first law of black hole mechanics,
\begin{equation}\label{eq:first_law}
\diff E=\beta^{-1} \diff S+ \Phi_p \,\diff Q_p + \Phi_w \,\diff Q_w\, ,
\end{equation}
is obeyed. Second, the corrections agree with those of \cite{Cano:2022tmn}, where three-charge black holes were considered, in the limit in which the third charge, associated to the presence of NS5 branes, goes  to zero. Finally, the expressions are consistent with T-duality, which exchanges $\gamma_p \leftrightarrow \gamma_w$ and sends $k_{\infty}\to 1/k_{\infty}$. One can see that the mass, entropy and temperature are left invariant, whereas the chemical potentials and charges are interchanged, as expected. 

\textbf{Four-dimensional black holes.} The expressions for the charges and mass read,
\begin{eqnarray}
	Q_{p}&=& \frac{\epsilon_p k_{\infty} }{8G_N} \rho_s \sinh \left(2\gamma_p\right)\,, \\[1mm]
	Q_{w}&=& \frac{\epsilon_{w} k_{\infty}^{-1} }{8G_N } \rho_s \sinh \left(2\gamma_w\right)\,, \\[1mm]
	E&=&\frac{\rho _s}{8G_N} \left[2+\cosh \left(2 \gamma _p\right)+\cosh \left(2 \gamma
	_w\right)\right]\, .
\end{eqnarray}
In turn, the inverse temperature and the chemical potential receive the following $\alpha'$ corrections,
\begin{eqnarray}
\beta&=&4\pi  \cosh \gamma_p \cosh\gamma_w \rho_s \left(1+\frac{\alpha' \Delta\beta}{\rho_s^2}\right)\,,\\[1mm]
	\Phi_{p}&=& \epsilon_p k_\infty^{-1} \tanh\gamma_p\left(1+\frac{\alpha' \Delta \Phi_{p}}{\rho_s^2}\right)
	\,,\\[1mm]
\Phi_{w}&=& \epsilon_w k_{\infty}\tanh\gamma_w\left(1+\frac{\alpha' \Delta \Phi_{w}}{\rho_s^2}\right)\,,
\end{eqnarray}
where
\begin{eqnarray}
\Delta \beta &=\,& \frac{\cosh (2\gamma_p)\left[1-2\cosh(2\gamma_w)\right]+\cosh (2\gamma_w)}{2\cosh (2\gamma_p)\left[1+2\cosh(2\gamma_w)\right]+2 \cosh (2\gamma_w)} - \frac{\epsilon_p \epsilon_w  \tanh \gamma_p \tanh \gamma_w}{8}   \,, \\[1mm]
\Delta\Phi_{p}&=\,& -\frac{2 \cosh (2\gamma_w)}{\cosh (2\gamma_p)\left[1+2\cosh(2\gamma_w)\right]+ \cosh (2\gamma_w)}- \frac{\epsilon_p \epsilon_w\tanh (\gamma_w)}{4\sinh (2\gamma_p)} \,, \\[1mm]
\Delta\Phi_{w}&=\,&  -\frac{2 \cosh (2\gamma_p)}{\cosh (2\gamma_p)\left[1+2\cosh(2\gamma_w)\right]+2 \cosh (2\gamma_w)}-\frac{\epsilon_p \epsilon_w\tanh (\gamma_p)}{4\sinh (2\gamma_w)} \,.
\end{eqnarray}
Finally, the expression for the entropy is
\begin{equation}
S = \frac{\pi \rho_s^2 \cosh \gamma_p \cosh \gamma_w}{G_N}  \left[1+ \frac{\alpha'}{2\rho_s^2}\left(1+ \frac{\epsilon_p \epsilon_{w}\tanh \gamma_p \tanh \gamma_w}{4} \right) \right]\,.
\end{equation}
These corrections agree with those of \cite{Zatti:2023oiq}, where the corrections to a family of four-charge black holes have been computed. As in the five-dimensional case, the thermodynamic quantities we have obtained transform as expected under T-duality and obey the first law of black hole mechanics \eqref{eq:first_law}.

\subsection{Thermodynamic quantities in the grand-canonical ensemble} 

In order to obtain the thermodynamics in the grand-canonical ensemble, we must consider a different choice of boundary conditions. This can be simply implemented by considering a different parametrization of the solution,
\begin{equation} \label{eqChangeEns}
\rho_s\to \rho_s+ \alpha' \delta \rho_s\left(\rho_s, \gamma_i\right)\,, \hspace{1cm} \gamma_{i}\to \gamma_{i}+\frac{\alpha' \delta \gamma_{i}\left(\rho_s, \gamma_j\right)}{\rho_s} \,, 
\end{equation}
and fixing $\delta \rho_s$ and $\delta \gamma_i$ by imposing the vanishing of the corrections to $\beta$ and the chemical potentials $\Phi_{i}$. The resulting expressions for the thermodynamic quantities associated to the five- and four-dimensional solutions are given below.

\textbf{Five-dimensional black holes:}
\begin{eqnarray}
\label{eq:chemicalpot_GCensemble}
\beta&=\,&2\pi \rho_s\cosh \gamma_p\, \cosh \gamma_w\,,\hspace{5mm}\Phi_{p}=\,\frac{\epsilon_{p} \tanh \gamma_p}{k_{\infty}}\,,\hspace{5mm}\Phi_{w}=\,\epsilon_{w}k_{\infty}\tanh \gamma_w\,,\\[1mm]
\label{eq:Qp}
Q_{p}&=\,&\frac{\epsilon_{p} k_{\infty}\pi}{8G_N}\left[\rho_s^2\sinh\left(2\gamma_p\right)+\alpha' \epsilon_{p}\epsilon_{w}\tanh \gamma_w\right]\,,\\[1mm]
\label{eq:Qw}
Q_{w}&=\,&\frac{\epsilon_{w}\pi}{8G_Nk_{\infty}}\left[\rho_s^2\sinh\left(2\gamma_w\right)+\alpha' \epsilon_{p}\epsilon_{w}\tanh \gamma_p\right]\,,\\[1mm]
\label{eq:E}
E&=\,&\frac{\pi \rho _s^2}{8G_N} \left[1+\cosh \left(2 \gamma _p\right)+\cosh \left(2 \gamma
   _w\right)+\frac{\alpha'}{4\rho_s^2} \left(-9+4 \epsilon_{p}\epsilon_{w}\tanh \gamma_p \tanh \gamma_w\right)\right]\,,\\[1mm]
\label{eq:S}
S&=\,& \frac{\pi^2\rho_s^3 \cosh \gamma_p \cosh \gamma_w}{2G_N}\,.
\end{eqnarray}

\textbf{Four-dimensional black holes:}
\begin{eqnarray}
\beta &= & 4 \pi \rho_s \cosh \gamma_p \cosh \gamma_w \,, \hspace{5mm}\Phi_p =  \epsilon_p k_\infty^{-1} \tanh \gamma_p \,, \hspace{5mm}\Phi_w = \epsilon_w k_\infty \tanh \gamma_w \,, \\[1mm]
Q_{p}& = & \frac{\epsilon_p k_{\infty}\rho_s \sinh \left(2\gamma_p\right) }{8G_N} \left[1+\frac{\alpha'}{2\rho_s^2}\left(1 + \frac{\epsilon_p \epsilon_{w} \tanh \gamma_w}{4 \tanh \gamma_p }\right)\right]\,, \label{eq:Qp4d} \\[1mm]
Q_{w}&= & \frac{\epsilon_{w} k_{\infty}^{-1} \rho_s \sinh \left(2\gamma_w\right)}{8G_N} \left[1+\frac{\alpha'}{2\rho_s^2}\left(1 + \frac{\epsilon_p \epsilon_{w} \tanh \gamma_p}{4 \tanh \gamma_w }\right)\right] \label{eq:Qw4d} \,,\\[1mm]
E\;  &= & \; \;\frac{\rho _s \left(\cosh \left(2 \gamma _p\right)+\cosh \left(2 \gamma_w\right)+2\right)}{8G_N} \left[1+\frac{\alpha'}{2\rho_s^2}\left(\frac{\cosh (2\gamma_p)+\cosh (2\gamma_w)-2 }{\cosh (2\gamma_p)+\cosh (2\gamma_w)+ 2} \right.\right.\nonumber\\[1mm]
&&\left.\left.+ \frac{\epsilon_p \epsilon_{w}\tanh \gamma_p \tanh \gamma_w}{4}\right)\right]\,, \label{eq:E4d} \\[1mm]
S  &=&  \frac{\pi \rho_s^2 \cosh \gamma_p \cosh \gamma_w }{G_N} \left[1+ \frac{\alpha'}{2\rho_s^2}\left(1 + \frac{\epsilon_p \epsilon_{w}\tanh \gamma_p \tanh \gamma_w}{4} \right) \right]  \label{eq:S4d} \,.
\end{eqnarray}

\section{Thermodynamics from the Euclidean on-shell action}\label{sec:chargesfromI}

In the saddle-point approximation the Euclidean on-shell action of the black hole gives the dominant contribution to the grand-canonical partition function \cite{Gibbons:1976ue}. This leads to the so-called quantum statisical relation,
\begin{equation}\label{eq:qsr}
I_{\infty}=\beta \,{\cal G}=\beta \left(E-\Phi_p \,Q_p-\Phi_w \,Q_w\right)-S\,,
\end{equation}
where $I_{\infty}$ is the renormalized Euclidean on-shell action and ${\cal G}$ is the grand-canonical potential, which is regarded as a function of the (inverse) temperature and the chemical potentials. Knowing ${\cal G}={\cal G}\left(\beta, \Phi_p, \Phi_w\right)$ suffices to extract all the thermodynamic quantities since the mass, charges and entropy can be obtained (assuming the first law of black hole mechanics) as follows:
\begin{equation}\label{eq:chargesfromI}
Q_{p}=-\frac{\partial {\cal G}}{\partial {\Phi}_{p}}\,, \hspace{5mm}Q_{w}=-\frac{\partial {\cal G}}{\partial {\Phi}_{w}}\,,\hspace{5mm} S=-\frac{\partial {\cal G}}{\partial{\beta^{-1}}}\, , \hspace{5mm} E={\cal G}+\Phi_p \,Q_p+\Phi_w \,Q_w+\beta^{-1} S\, .
\end{equation}
As shown e.g. in \cite{Reall:2019sah, Bobev:2022bjm, Cassani:2022lrk}, this method to obtain the thermodynamics is particularly useful when dealing with higher-derivative corrections.

The purpose of this section is to evaluate the Euclidean on-shell action of the two-charge black holes at first order in $\alpha'$ and check that the thermodynamics that we get match the ones obtained in the previous section. For simplicity, we are going to evaluate the $(d+1)$-dimensional Euclidean effective action in the string frame, since its dimensional reduction on ${\mathbb S}^1_{y}$ gives rise to much more terms \cite{Baron:2017dvb, Elgood:2020xwu, Ortin:2020xdm, Eloy:2020dko, Liu:2023fqq}. Instead, the $(d+1)$-dimensional action coincides with the ten-dimensional one \eqref{eq:actionheterotic} up to an overall factor which is absorbed in the $(d+1)$-dimensional Newton constant ${\hat G}_{N}$.  

Then, the heterotic Euclidean on-shell action $I$ for a manifold $M$ with boundary $\partial M$ is given by 
\begin{equation}\label{eq:Euclidean_action}
I=-\frac{{\hat g}_s^2}{16\pi {\hat G}_N}\int_{M} \diff^{d+1}x \sqrt{|\hat g|} \, {\cal L}_{\rm {eff}}+\frac{{\hat g}_s^2}{8\pi {\hat G}_N}\int_{\partial M}\diff^{d}x \sqrt{|\hat h|}\, e^{-2\hat \phi}{\hat K}+\dots\,,
\end{equation}
where 
\begin{equation}
 {\cal L}_{\rm {eff}}=e^{-2\hat \phi}\left [\hat R-4\,\partial^{\hat\mu}{\hat \phi}\,\partial_{\hat\mu} \hat \phi+\frac{1}{2\cdot 3!}{\hat H}^2+\frac{\alpha'}{8}{\hat R}_{(-)}{}_{\hat \mu \hat \nu \hat a \hat b}{\hat R}_{(-)}{}^{\hat\mu \hat\nu \hat a \hat b}\right] \,
\end{equation}
is the effective Lagrangian of the heterotic superstring at first order in $\alpha'$ (see appendix~\ref{app:eff_action+EOMs}). The second term in \eqref{eq:Euclidean_action} is the standard Gibbons-Hawking-York (GHY) term written in the string frame: ${\hat h}_{\mu\nu}$ represents the metric induced at $\partial M$ and $\hat K$ is the trace of the extrinsic curvature. Finally, the dots indicate additional boundary terms associated to the higher-derivative corrections, which on general grounds are expected to give a vanishing contribution for asymptotically-flat solutions, \cite{Reall:2019sah} (hence, we shall ignore them from now on). As observed in \cite{Tseytlin:1988tv, Chen:2021dsw}, the bulk contribution reduces to a boundary term after using the equation of motion of the dilaton \eqref{eq:dilatoneom}, which implies
\begin{equation}
{\cal L}_{\rm{eff}}=-2{\hat \nabla}^2 e^{-2\hat \phi}\, .
\end{equation}
Therefore, we have that \eqref{eq:Euclidean_action} reduces to:
\begin{equation}\label{eq:I}
I=\frac{{\hat g}^2_s}{8\pi {\hat G}_N}\int_{\partial M}\diff^{d}x \sqrt{|\hat h|}\, e^{-2\hat \phi}\left({\hat K}-2\,n^{\hat \mu}\, \partial_{\hat \mu}\hat \phi\right)\,,
\end{equation}
where $n^{\hat\mu}$ is the unit normal to the boundary. 

Here we are interested in asymptotically-flat black holes whose boundary ${\partial M}$ has the topology of ${\mathbb S}^1_\beta \times {\mathbb S}^{d-2}$ ($\times {\mathbb S}^1_y$). As it is well known, the GHY term diverges in the limit in which the radius of the ${\mathbb S}^{d-2}$ goes to infinity, just as in flat spacetime. In order to obtain a finite on-shell action, we follow the prescription of \cite{Gibbons:1976ue}. This amounts to first consider a regulated spacetime $M_R$, where $R$ is a radial cutoff. The regulated spacetime then corresponds to the region $\rho\le R$, and its boundary $\partial M_{R}$ is the hypersurface $\rho=R$. Second, we introduce an auxiliary configuration with flat metric ${\hat\delta}_{R}$ and constant dilaton ${\hat \phi}_{R}$ chosen so that the induced fields (metric and dilaton) at $\rho=R$ coincide with the induced metric and dilaton of the black hole solution, namely ${\hat \delta}_{R}|_{\rho=R}={\hat g}|_{\rho=R}$ and $\hat \phi_R=\hat \phi|_{\rho=R}$. Once we have $\hat \delta_R$ and $\hat \phi_R$, we substract the regulated action associated with the flat spacetime $I_R[\hat\delta_R, \hat \phi_R]$ to the one associated with the black hole $I_R[\hat g, \hat \phi]$  and only then take the $R\to \infty$ limit. Summarizing, the renormalized action $I_{\infty}$ is given by
\begin{equation}
I_{\infty}=\lim_{R\to \infty}\left(I_R[\hat g, \hat \phi]-I_R[\hat\delta_R, \hat \phi_R]\right)\, ,
\end{equation}
and, making use of \eqref{eq:I}, we get
\begin{equation}\label{eq:I_infty}
I_{\infty}=\lim_{R\to \infty}\left\{\frac{{\hat g}_s^2}{8\pi {\hat G}_N}\int_{\rho=R}\diff^{d}x \sqrt{|\hat h|}\, e^{-2\hat \phi}\left[\left({\hat K}-{\hat K}_{{\hat\delta}_ R}\right)-2\,n^{\hat \mu}\, \partial_{\hat \mu}\hat \phi\right]\right\}\,,
\end{equation}
where ${\hat K}_{{\hat\delta}_ R}$ is the trace of the extrinsic curvature associated to the metric ${\hat\delta}_{R}$. For the two-charge black holes we are interested in, the auxiliary flat solution $\{{\hat \delta}_R, \hat \phi_R\}$ is given by
\begin{equation}
\begin{aligned}
-\hat \delta_R=\,&\frac{f(R)}{\zp(R)  {\tilde f}_{w}(R)}\diff\tau^2+\zz(R)\left(\diff\rho^2+\rho^2 \diff\Omega^2_{(d-2)}\right)+k^2_{R}\left[\diff y+\beta_{p}k^{-1}_{\infty}\left(\zp(R)^{-1}-1\right)\diff t\right]^2\,,\\
\hat \phi_R=\,&\hat \phi (R)\,,
\end{aligned}
\end{equation}
where $k_{R}^2=k^2_{\infty}\frac{\zp (R)}{\zm (R)}$. Now we have all the ingredients to evaluate \eqref{eq:I_infty} using the corrected solutions found in the previous section. Let us do this for the five- and four-dimensional solutions separately.

\textbf{Five-dimensional black holes.} Expressing the result in the grand-canonical ensemble, we get that the Euclidean on-shell action of the five-dimensional two-charge black holes is given by
\begin{equation}\label{eq:onshellaction5d}
I_{\infty}=\frac{\pi^2 {\rho}_s^3\cosh \gamma_p\cosh\gamma_w}{4 G_N}\left[1-\frac{9\alpha'}{4{\rho}_s^2}-\frac{\alpha' \epsilon_{p}\epsilon_w\tanh\gamma_p\tanh\gamma_w}{{ \rho}^2_s}\right]\, ,
\end{equation}
and we recall that
\begin{equation}
\beta=2\pi \rho_s\cosh \gamma_p\, \cosh \gamma_w\,,\hspace{5mm}\Phi_{p}=\,\frac{\epsilon_{p}\tanh \gamma_p}{k_{\infty}}\,,\hspace{5mm}\Phi_{w}=\,\epsilon_{w}k_{\infty}\tanh \gamma_w\,.
\end{equation}
It is a straightforward calculation to show that the corrected charges that follow from the on-shell action (using \eqref{eq:chargesfromI}) are in perfect agreement with the ones we computed in the previous section, namely with eqs.~\eqref{eq:Qp}, \eqref{eq:Qw}, \eqref{eq:E} and \eqref{eq:S}.

\textbf{Four-dimensional black holes.} In the four-dimensional case the on-shell action in the grand-canonical ensemble takes the form
\begin{equation}\label{eq:onshellaction4d}
I_{\infty} = \frac{\pi \rho_s^2 \cosh \gamma_p \cosh \gamma_w}{G_N}  \left[1- \frac{\alpha'}{2\rho_s^2}\left(1 + \frac{\epsilon_p \epsilon_{w}\tanh \gamma_p \tanh \gamma_w}{4} \right) \right]  \,,
\end{equation}
with the inverse temperature and the chemical potentials given by
\begin{equation}
	\beta= 4\pi \rho_s\cosh \gamma_p\, \cosh \gamma_w\,,\hspace{5mm}\Phi_{p}=\,\frac{\epsilon_p \tanh \gamma_p}{k_\infty}\,,\hspace{5mm}\Phi_{w}=\,\epsilon_{w}\,k_{\infty}\tanh \gamma_w\,.
\end{equation}
As before, the charges (\ref{eq:Qp4d}), (\ref{eq:Qw4d}), (\ref{eq:E4d}) and (\ref{eq:S4d}) are properly recovered from \eqref{eq:chargesfromI}.


\section{Two-charge black holes from Schwarzshchild-Tangherlini}\label{sec:2QsBHfromST}

As already mentioned, the corrections to the thermodynamics of two-charge black holes have been previously studied in \cite{Giveon:2009da, Chen:2021dsw}. The strategy of these references is to find the $\alpha'$ corrections by performing a set of $O(2, 2)$ transformations (boost with parameter $\delta_{w}$ plus T-duality along $y$, followed by another boost with parameter $\delta_{p}$) to the Schwarzschild-Tangherlini black hole, whose $\alpha'$ corrections had been already studied in \cite{Callan:1988hs}. The main difference between these two references is that \cite{Chen:2021dsw} just focuses on the thermodynamic properties while in \cite{Giveon:2009da} the full corrected solutions are obtained by means of this technique. This is technically more complicated than just obtaining the thermodynamics, as one has to take into account the explicit $\alpha'$ corrections to the $O(2, 2)$ transformations. This might be the reason why the $\alpha'$-corrected thermodynamics obtained in these references do not agree with one another.

The goal of this section is to show that our results for the $\alpha'$-corrected thermodynamics of heterotic two-charge black holes are in agreement with those of \cite{Chen:2021dsw}. To this aim, we find convenient to review here their calculation. A key observation is that the Euclidean on-shell action remains invariant after the $O(2,2)$ transformation. Therefore,
\begin{equation}\label{eq:I=tildeI}
I_{\infty}(\beta, \Phi_p, \Phi_w; \phi_{\infty}, k_{\infty})={\tilde I}_{\infty}(\tilde\beta; {\tilde\phi}_{\infty}, {\tilde k}_{\infty})\, ,
\end{equation}
where, following the conventions of \cite{Chen:2021dsw}, we are using tildes for the quantities associated to the Schwarzschild-Tangherlini solution.

The right-hand side of \eqref{eq:I=tildeI} is obtained from the $\alpha'$ corrections to the Schwarzschild-Tangherlini solution \cite{Callan:1988hs}. Focusing just on the thermodynamic quantities, we have 
\begin{equation}
{\tilde E}=\frac{d-2}{d-3}\frac{\gamma_d \,{{\tilde R}_{\beta}}^{d-3}}{8\pi {\tilde G}_{N}} \left(1-\frac{\epsilon_d \, \alpha'}{4 {\tilde R}_{\beta}^2}\right)\,, \hspace{5mm} {\tilde S}= \frac{\gamma_d \, {\tilde R}_{\beta}^{d-2}}{4{\tilde G}_{N}}\left(1-\frac{\sigma_d \, \alpha'}{4 {\tilde R}_{\beta}^2}\right)\,,
\end{equation}
where ${\tilde R}_{\beta}\equiv {\tilde\beta}/(2\pi)$ is the  radius of the thermal circle ${\mathbb S}^1_{\beta}$ and
\begin{equation}
\gamma_d=\omega_{d-2} \left(\frac{d-3}{2}\right)^{d-2}\,, \hspace{5mm}\epsilon_d=\frac{2(d-4)(d-2)}{d-3}\,, \hspace{5mm}\sigma_d=\frac{2(d-5)(d-2)^2}{(d-3)^2}\, .
\end{equation}
Assuming the quantum statistical relation \eqref{eq:qsr}, we get that the Euclidean on-shell action of the Schwarzschild-Tangherlini black hole is
\begin{equation}\label{eq:I_ST}
{\tilde I}_{\infty}={\tilde \beta} {\tilde E}- {\tilde S}= \frac{\gamma_d \, {\tilde R}_{\beta}^{d-2}}{4{\tilde G}_{N} (d-3)} \left[1-\frac{(d-2)^2\alpha'}{2(d-3){\tilde R}_{\beta}^{2}}\right]\,.
\end{equation}
Because of \eqref{eq:I=tildeI}, the right-hand side of  \eqref{eq:I_ST} computes the Euclidean on-shell action of the two-charge black holes as well. This is however meaningless at this stage, since we have not yet specified the expressions for $\beta$ and the chemical potentials $\Phi_{p}$, $\Phi_{w}$ in terms of ${\tilde R}_{\beta}$ and the parameters of the $O(2,2)$ transformations. Such expressions can be recovered from \cite{Chen:2021dsw}. Taking into account all the possibilities for the signs of the winding and momentum charges, we find
\begin{eqnarray}
\label{eq:relation_beta}
R_{\beta}&=\,& {\tilde R}_{\beta}\cosh \delta_p\, \cosh \delta_w \left(1-\frac{\alpha' \epsilon_{p}\epsilon_{w}\tanh \delta_p \tanh \delta_w}{2{\tilde R}_{\beta}^2}\right)\,,\\[1mm]
\label{eq:relation_Phip}
\Phi_{p}&=\,&\frac{\epsilon_{p}\tanh \delta_p}{k_{\infty}}\left(1-\frac{\alpha' \epsilon_{p}\epsilon_{w}\tanh \delta_w}{{\tilde R}_{\beta}^2\sinh \left(2\delta_p\right)}\right)\,,\\[1mm]
\label{eq:relation_Phiw}
\Phi_{w}&=\,&\epsilon_{w} k_{\infty}\tanh \delta_w\left(1-\frac{\alpha' \epsilon_{p}\epsilon_{w}\tanh \delta_p}{{\tilde R}_{\beta}^2\sinh \left(2\delta_w\right)}\right)\,,
\end{eqnarray}
where $\delta_{p, w}$ represent the parameters of the $O(2, 2)$ transformations. In addition to this, one must also bear in mind the relation between the moduli of the solutions. In particular, we need the relation between the asymptotic values of the $d$-dimensional dilaton $e^{\phi_{\infty}}=g_s$, which is the following \cite{Chen:2021dsw}
\begin{equation}\label{eq:relation_gs}
g_s^2=\,{\tilde g}_s^2 \cosh \delta_p\, \cosh \delta_w \left(1-\frac{\alpha' \epsilon_{p}\epsilon_{w}\tanh \delta_p \tanh \delta_w}{2{\tilde R}_{\beta}^2}\right)\,.
\end{equation}
Taking into account that $G_N\propto g_s^2$, one gets that the Newton constants are related by
\begin{equation}
G_N={\tilde G}_{N}\cosh \delta_p\, \cosh \delta_w \left(1-\frac{\alpha' \epsilon_{p}\epsilon_{w}\tanh \delta_p \tanh \delta_w}{2{\tilde R}_{\beta}^2}\right)\, .
\end{equation}
Using this in \eqref{eq:I_ST}, we obtain
\begin{equation}
I_{\infty}={\tilde \beta} {\tilde E}- {\tilde S}= \frac{\gamma_d \, {\tilde R}_{\beta}^{d-2}\cosh \delta_p \cosh \delta_w}{4{G}_{N} (d-3)} \left[1-\frac{\alpha'}{2 {\tilde R}_{\beta}^2}\left(\frac{(d-2)^2}{(d-3)}+\epsilon_{p}\epsilon_w \tanh \delta_p \tanh \delta_w\right)\right]\,.
\end{equation}
This already specifies the thermodynamics. However, the parametrization we are using here differs from the one(s) used in the previous sections. It is not difficult to find that the relation between ${\tilde R}_{\beta}, \delta_p, \delta_w$ and the parameters $\rho_s, \gamma_p, \gamma_w$ used in the previous sections to express the thermodynamics in the grand-canonical ensemble is given by 
\begin{eqnarray}
{\tilde R}_{\beta}&=&\frac{2\rho_s}{d-3} \left(1-\frac{\epsilon_{p}\epsilon_{w} (d-3)^2\alpha'\tanh \gamma_p \tanh \gamma_w}{8\rho_s^2}\right)\, ,\\[1mm]
\delta_p&=&\gamma_p+\frac{\epsilon_{p}\epsilon_{w} (d-3)^2\alpha' \tanh \gamma_w}{8\rho_s^2}\,,\\[1mm]
\delta_w&=&\gamma_w+\frac{\epsilon_{p}\epsilon_{w} (d-3)^2\alpha' \tanh \gamma_p}{8\rho_s^2}\,.
\end{eqnarray}
Making use of these relations, we can write the on-shell action of the two-charge black holes in the grand-canonical ensemble:
\begin{equation}
I_{\infty}=\frac{\omega_{d-2} \,\rho_s^{d-2}\, \cosh \gamma_p \cosh \gamma_w}{4 (d-3)G_N}\left\{1-\frac{(d-3)\alpha'}{8\rho_s^2}\left[(d-2)^2+\epsilon_{p}\epsilon_{w} (d-3)^2\tanh\gamma_p\tanh \gamma_w\right]\right\}\, .
\end{equation}
This properly reduces to \eqref{eq:onshellaction5d} and to \eqref{eq:onshellaction4d} when setting $d=5$ and $d=4$, respectively. Given the grand-canonical potential ${{\cal G}=\beta^{-1}I_{\infty}}$, we can obtain the charges, entropy and mass through \eqref{eq:chargesfromI}, as already discussed.
Expressing them in the grand-canonical ensemble, we obtain the following expressions
\begin{eqnarray}
\beta&=\,&\frac{4\pi \rho_s}{d-3}\cosh \gamma_p\, \cosh \gamma_w\,,\hspace{5mm}
\Phi_{p}=\,\frac{\epsilon_p\tanh \gamma_p}{k_{\infty}}\,,\hspace{5mm}
\Phi_{w}=\,{\epsilon}_{w}k_{\infty}\tanh \gamma_w\,,\\[1mm]
Q_p&=& Q^{(0)}_{p}\left[1-\frac{(d-3)^2\alpha'}{16\rho_s^2}\left(\sigma_d-2\epsilon_p\epsilon_{w}(4-d+\text{coth}^2\gamma_p)\tanh\gamma_p\tanh \gamma_w\right)\right]\,,\\[1mm]
Q_w&=& Q^{(0)}_{w}\left[1-\frac{(d-3)^2\alpha'}{16\rho_s^2}\left(\sigma_d-2\epsilon_p\epsilon_{w}(4-d+\text{coth}^2\gamma_w)\tanh\gamma_p\tanh \gamma_w\right)\right]\,,\\[1mm]
S&=&S^{(0)}\left[1-\frac{(d-3)^2\alpha'}{16\rho_s^2}\left(\sigma_d+2(d-5)\epsilon_p\epsilon_{w}\tanh\gamma_p\tanh\gamma_w\right)\right]\, ,
\end{eqnarray}
where 
\begin{equation}
\begin{aligned}
Q^{(0)}_p=\,&\frac{(d-3)\epsilon_pk_{\infty}\omega_{d-2}\rho_s^{d-3}\sinh (2\gamma_p)}{32\pi G_N}\,, \hspace{5mm} Q^{(0)}_w=\,\frac{(d-3)\epsilon_{w}\omega_{d-2}\rho_s^{d-3}\sinh (2\gamma_w)}{32\pi G_N k_{\infty}}\,, \\[1mm] 
S^{(0)}=\,&\frac{\gamma_{d-2}\rho_s^{d-2}\cosh \gamma_p \cosh \gamma_w}{32G_N}\, .
\end{aligned}
\end{equation}
Instead of the mass we provide the expression for the grand-canonical potential ${\cal G}$, which is simpler
\begin{equation}
{\cal G}=\frac{\omega_{d-2} \, \rho_s^{d-3}}{16\pi G_N} \left[1-\frac{(d-3)\alpha'}{8\rho_s^2}\left((d-2)^2+(d-3)^2\,\epsilon_p\epsilon_{w}\tanh\gamma_p\tanh\gamma_w\right)\right]\,.
\end{equation}
The mass $E$ follows then from the last of \eqref{eq:chargesfromI}. It is now straightforward to compare these expressions with the ones we obtained in sections~\ref{sec:BH_thermodynamics} and \ref{sec:chargesfromI} and see that they are in perfect agreement. Furthermore, we have also checked that they agree with the corrected thermodynamics given in the appendix of \cite{Chen:2021dsw}, after using the map between the two parametrizations, provided in \eqref{eq:relation_beta}, \eqref{eq:relation_Phip} and \eqref{eq:relation_Phiw}.

\section*{Acknowledgements}
We are grateful to Pablo A. Cano, Tom\'as Ort\'in and Pedro F. Ram\'irez for many discussions and collaborations along these years on this and related topics. We would like to further thank the FISPAC group at University of Murcia and the group in Hamburg for their feedback while presenting this work and for hospitality. AR is supported by a postdoctoral fellowship associated to the MIUR-PRIN contract 2020KR4KN2, ``String Theory as a bridge between Gauge Theories and Quantum Gravity''. M.Z. has been supported by the fellowship LCF/BQ/DI20/11780035 from ``La Caixa'' Foundation (ID100010434), by the MCI, AEI, FEDER (UE) grants PID2021-125700NBC21 (“Gravity, Supergravity and Superstrings” (GRASS)) and PID2021-123017NB-I00, and by the grant IFT Centro de Excelencia Severo Ochoa CEX2020-001007-S.

\appendix

\section{Effective action and equations of motion}\label{app:eff_action+EOMs}
The effective action of the heterotic string at first order in $\alpha'$ in the Bergshoeff-de Roo formulation \cite{Bergshoeff:1989de} is given by,
\begin{equation}\label{eq:actionheterotic}
\begin{aligned}
S_{\rm{eff}}=\int \diff^{10}x \sqrt{-{\hat g}}\,e^{-2\hat \phi}\left [\hat R-4\,\partial^{\hat\mu}{\hat \phi}\,\partial_{\hat\mu} \hat \phi+\frac{1}{2\cdot 3!}{\hat H}^2+\frac{\alpha'}{8}{\hat R}_{(-)}{}_{\hat\mu \hat\nu \hat a\hat b}{\hat R}_{(-)}{}^{\hat\mu \hat\nu \hat a\hat b}\right]\, ,
\end{aligned}
\end{equation}
where ${\hat R}_{(-)}{}^{\hat a}{}_{\hat b}=\frac{1}{2}{\hat R}_{(-)}{}_{\hat\mu \hat\nu}{}^{\hat a}{}_{\hat b} \, dx^{\hat \mu}\wedge dx^{\hat \nu}$ is defined as the curvature two-form associated to the torsionful spin connection,
\begin{equation}\label{eq:Rminus}
{\hat R}_{(-)}{}^{\hat a}{}_{\hat b}=\diff{\hat\omega}_{(-)}{}^{\hat a}{}_{\hat b}-{\hat\omega}_{(-)}{}^{\hat a}{}_{\hat c}\wedge {\hat\omega}_{(-)}{}^{\hat c}{}_{\hat b}\, ,
\end{equation}
where
\begin{equation}\label{eq:torsion}
{\hat\omega}_{(-)}{}^{\hat a}{}_{\hat b}={\hat\omega}^{\hat a}{}_{\hat b}-\frac{1}{2}{\hat H}_{c}{}^{\hat a}{}_{\hat b}\,{\hat e}^{\hat c}\, .
\end{equation}
In the formula above, ${\hat \omega}^{\hat a}{}_{\hat b}$ represents the standard Levi-Civita spin connection and the torsion is determined by the three-form field strength ${\hat H}$, which satisfies the modified Bianchi identity \cite{Green:1984sg},
\begin{equation}
\label{eq:bianchi}
\diff{\hat H}-\frac{\alpha'}{4}{\hat R}_{(-)}{}^{\hat a}{}_{\hat b}\wedge {\hat R}_{(-)}{}^{\hat b}{}_{\hat a}=0\, .
\end{equation} 
This implies that, locally, $\hat H$ is given by
\begin{equation}\label{eq:defH}
{\hat H}=\diff{\hat B}+\frac{\alpha'}{4}{\hat \Omega}_{(-)}\, ,
\end{equation}
where ${\hat\Omega}_{(-)}$ is the Lorentz Chern-Simons three-form, defined as
\begin{equation}
\label{eq:cs3form}
{\hat \Omega}_{(-)}=\diff{\hat \omega}_{(-)}{}^{\hat a}{}_{\hat b}\wedge {\hat \omega}_{(-)}{}^{\hat b}{}_{\hat a}-\frac{2}{3} \,{\hat \omega}_{(-)}{}^{\hat a}{}_{\hat b}\wedge {\hat \omega}_{(-)}{}^{\hat b}{}_{\hat c}\wedge {\hat \omega}_{(-)}{}^{\hat c}{}_{\hat a}\, .
\end{equation}
Let us note that the definition of the Kalb-Ramond two-form $\hat B$ in \eqref{eq:defH} is a recursive one, since $\hat \Omega_{(-)}$ depends on $\hat H$ through \eqref{eq:torsion}. Then, it should be implemented order by order in $\alpha'$, as follows:
\begin{equation}
{\hat H}^{(0)}=\diff {\hat B}^{(0)}\,, \hspace{5mm} {\hat H}^{(1)}=\diff{\hat B}^{(1)}+\frac{\alpha'}{4}{\hat \Omega}^{(0)}_{(-)}\, ,\hspace{5mm} ...  \hspace{5mm}{\hat H}^{(n)}=\diff{\hat B}^{(n)}+\frac{\alpha'}{4}{\hat \Omega}^{(n-1)}_{(-)}\, ,
\end{equation}
where ${\hat \Omega}^{(n-1)}_{(-)}$ is computed using ${\hat H}^{(n-1)}$. This implies that the term ${\hat H}^2$ in \eqref{eq:actionheterotic} actually contains and infinite tower of $\alpha'$ corrections.  By consistency, we only keep the first-order ones.

\subsection{Equations of motion}

The derivation of the equations of motion drastically simplifies when using a lemma proven in \cite{Bergshoeff:1989de}. It states that the variation of the action with respect to ${\hat\omega}_{(-)}{}^{\hat a}{}_{\hat b}$ produces terms which are subleading in $\alpha'$. Hence, we only need to vary the terms where the fieds appear explicitly, ignoring implicit ocurrences of the fields through ${\hat\omega}_{(-)}{}^{\hat a}{}_{\hat b}$. This leads to the following equations of motion,
\begin{eqnarray}
\label{eq:einsteineom}
{\hat R}_{\hat \mu\hat\nu}-2{\hat \nabla}_{\hat\mu}\partial_{\hat\nu}{\hat\phi}+\frac{1}{4}{\hat H}_{\hat\mu\hat\rho\hat\sigma}{\hat H}_{\hat\nu}{}^{\hat\rho\hat\sigma}+\frac{\alpha'}{4}{\hat R}_{(-)}{}_{\hat\mu\hat\rho \hat a\hat b}{\hat R}_{(-)}{}_{\hat\nu}{}^{\hat\rho \hat a\hat b}&=& {\cal O}{(\alpha'^2)}\, , \\[1mm]
\label{eq:dilatoneom}
2{\hat \nabla}^2 e^{-2\hat \phi}+e^{-2\hat \phi}\left [\hat R-4\,\partial^{\hat\mu}{\hat \phi}\,\partial_{\hat\mu} \hat \phi+\frac{1}{2\cdot 3!}{\hat H}^2+\frac{\alpha'}{8}{\hat R}_{(-)}{}_{\hat\mu \hat\nu \hat a \hat b}{\hat R}_{(-)}{}^{\hat\mu \hat\nu \hat a\hat b}\right]&=&{\mathcal O}\left(\alpha'^2\right)\, ,\\[1mm]
\label{eq:KReom}
\diff\left(e^{-2\hat \phi}\hat\star \hat H\right)&=&{\mathcal O}\left(\alpha'^2\right) .
\end{eqnarray}
As observed in \cite{Tseytlin:1988tv, Chen:2021dsw}, the equation of motion of the dilaton tells us that
\begin{equation}
{\cal L}_{\rm{eff}}=-2{\hat \nabla}^2 e^{-2\hat \phi}\, ,
\end{equation}
where ${\cal L}_{\rm{eff}}$ is the effective Lagrangian. This observation is really helpful to evaluate the Euclidean on-shell action, as it allows us to reduce the problem to the evaluation of a boundary term. 

\section{Additional details on the procedure  to find the $\alpha'$-corrected solutions}\label{app:correctedsol}
The purpose of this appendix is to explain in more detail the procedure we have followed to solve the $\alpha'$-corrected equations motion \eqref{eq:einsteineom}, \eqref{eq:dilatoneom} and \eqref{eq:KReom}.

\subsection*{The ansatz} 

Let us begin motivating the ansatz we have used for the metric $\hat g_{\mu\nu}$ and two-form $\hat B$. This is given in \eqref{eq:ansatz_metric} and \eqref{eq:ansatz_B}, which we repeat here for convenience:
\begin{eqnarray}
\diff{\hat s}^2&=&\frac{f}{\zp  {\tilde f}_{w}}\diff t^2-\zz\left(f^{-1}{\diff}\rho^2+\rho^2 {\diff}\Omega^2_{(d-2)}\right)-k^2_{\infty}\frac{\zp}{ {\tilde f}_{w}}\left[{\diff}y+\beta_{p}k^{-1}_{\infty}\left(\zp^{-1}-1\right){\diff}t\right]^2\,,\\[1mm]
{\hat B}&=&\beta_w k_{\infty}\left({f}_{w}^{-1}-1\right) {\diff}t \wedge {\diff}y\,.
\end{eqnarray}
For the dilaton we only assume a dependence on just the radial coordinate $\rho$.

The ``recipe'' followed to fix the ansatz is essentially to keep the same field components active as in the two-derivative solution. Spherical symmetry reduces the number of independent components of the metric to four: ${\hat g}_{tt}, {\hat g}_{\rho \rho}, {\hat g}_{yy}$ and ${\hat g}_{ty}$. These are in one-to-one correspondence with the functions $f, f_p, {\tilde f}_{w}$ and $g$. The reason to choose this particular parametrization is that we expect the form of these functions will be simpler, just by experience with the two-derivative ones. The last function to be considered is $f_{w}$, which is associated to the only non-vanishing component of the two-form $\hat B$. Since we are going to treat the $\alpha'$ corrections in a perturbative fashion, the form of these functions must be:
\begin{equation}
\begin{aligned}
{f}_{p}=\,&1+\frac{q_p}{\rho^{d-3}}+\alpha' {\delta f_p}\, , \hspace{5mm} {\tilde f}_{w}=\,1+\frac{q_w}{\rho^{d-3}}+\alpha' {\delta {\tilde f}_{w}}\, ,\hspace{5mm}\zz=\,1+\alpha' {\delta \zz}\, ,\\[1mm]
{f}=&\,1-\frac{\rho_s^2}{\rho^{d-3}}+\alpha' {\delta f}\, ,\hspace{5mm} {f}_{w}=\,1+\frac{q_w}{\rho^{d-3}}+\alpha' {\delta {f}_{w}}\, ,
\end{aligned}
\end{equation}
so that the two-derivative solution is properly recovered in the  $\alpha'\to 0$ limit. 

When plugging the above ansatz in the corrected equations of motion, one gets a coupled system of second-order differential equations for the unknown functions. In what follows we describe the procedure that we have followed in order to solve it, focusing on the five-dimensional case.

\subsection*{The equation of motion of $\hat B$} 

The expression for the dilaton can be found by solving the equation of motion of the two-form $\hat B$, \eqref{eq:KReom}. There is just one independent component which is not trivially satisfied, and it yields the following equation (for $d=5$):
\begin{equation}
\frac{f''_w}{f'_w}-\frac{2f'_w}{f_w}+\frac{g'}{g}+\frac{{\tilde f}'_w}{{\tilde f}'_w}+\frac{3}{\rho}-2\hat \phi'=0\,,
\end{equation}
where primes denote derivatives with respect to $\rho$. This equation can be integrated once to give 
\begin{equation}
\left[\hat \phi-\frac{1}{2}\log\frac{-\rho^3 g {\tilde f}_w f'_w}{f_w^2}\right]'=0\, ,
\end{equation}
which is solved by 
\begin{equation}
\hat \phi = a_{\phi} + \frac{1}{2}\log\frac{-\rho^3 g {\tilde f}_w f'_w}{2q_w f_w^2}\, ,
\end{equation}
where $a_{\phi}$ is an appropriate integration constant which we are going to fix imposing that the asymptotic value of the dilaton (string coupling) is not renormalized. When expressing the solution in the microcanonical ensemble, the term inside the logarithm goes to $1$ at infinity, which means that $a_{\phi}$ is identified with ${\hat\phi}_{\infty}$.

\subsection*{Einstein and dilaton equations}

As explained in the main text, the strategy to solve the corrected Einstein and dilaton equations is to expand the unknown functions $\Psi=\{\delta f, \delta f_p, \delta f_w, \delta {\tilde f}_w, \delta g\}$ in a series in $1/\rho^2$, 
\begin{equation}\label{eq:asymptotic_exp}
\Psi=\frac{a_{\Psi}}{\rho^2}+\sum_{n>1}^N \frac{b^{(n)}_{\Psi}}{\rho^{2n}}\,,
\end{equation}
and then solve \eqref{eq:einsteineom} and \eqref{eq:dilatoneom} order by order in $1/\rho^2$. This leads to a set of algebraic equations that determine the values of the coefficients $b_{\Psi}^{(n)}$ in terms of $a_{\Psi}$ and of the parameters of the two-derivative solution, $\rho_s, q_p, q_w$. Two out of the five integration constants $a_{\Psi}$ can be fixed right away. These are $a_{f_p}$ and $a_{f_w}$, which are both set to zero by imposing that the charges of the black hole $Q_p$ and $Q_w$ do not receive $\alpha'$ corrections. The next step is to find the generating functions that produce the asymptotic expansions \eqref{eq:asymptotic_exp}. This is done with the help of \texttt{Mathematica}.\footnote{For this purpose, one has to compute the solution up to sufficiently high order in $1/\rho^2$. }
Finally, we must fix the three remaining integration constants $a_g$, $a_g$ and $a_{\tilde f_p}$. Since we want to express the solution in the microcanonical ensemble, we must fix one of these constants (let us say, $a_g$) by imposing the mass does not receive $\alpha'$ corrections.  The resulting solution turns out to be singular at the horizon for arbitrary values of the two remaining integration constants, $a_f$ and $a_{\tilde f_p}$.\footnote{In particular, the dilaton and the Kaluza-Klein scalar diverge when $\rho \to \rho_H$.} Demanding regularity  imposes two conditions which fix both $a_f$ and $a_{\tilde f_p}$, leaving us with the solution reported in section~\ref{sec:alpha_corrected_BHs}.

\section{Dimensional reduction on a circle}\label{app:dimensional_red}

In this appendix we make use of the results of \cite{Elgood:2020xwu} in order to find the dimensional reduction of the solutions to $d$ dimensions. The $d$-dimensional fields are: the (string-frame) metric $g_{\mu\nu}$, the dilaton $\phi$, the Kaluza-Klein scalar $k$ and vector $A_{\mu}$, the two-form $B_{\mu\nu}$ and, finally, the winding vector $C_{\mu}$.\footnote{Note that in this section $\mu, \nu=0, \dots, d-1$, as opposed to the rest of the paper, where $\mu, \nu=0, \dots, d$.} These are given in terms of the higher-dimensional fields by the following expressions \cite{Elgood:2020xwu}:
\begin{eqnarray}
g_{\mu\nu}&=&{\hat g}_{\mu\nu}-\frac{{\hat g}_{\mu y}{\hat g}_{\nu y}}{{\hat g}_{yy}}\, , \hspace{1cm} A_{\mu}=\frac{{\hat g}_{\mu y}}{{\hat g}_{yy}}\,, \hspace{1cm} k^2=- {\hat g}_{yy}\,, \\[1mm]
B_{\mu\nu}&=&\, {\hat B}_{\mu\nu}+{\hat g}_{y[\mu}{\hat B}_{\nu]y}+\frac{\alpha'}{4}\,\frac{{\hat g}_{y [\mu}{\hat\Omega}_{(-)}{}_{\nu]}{}^{\hat a}{}_{\hat b}\,{\hat\Omega}_{(-)}{}_{y}{}^{\hat b}{}_{\hat a}}{{\hat g}_{yy}}\, , \\[1mm]
\label{eq:Cmu}
C_{\mu}&=&{\hat B}_{\mu y}-\frac{\alpha'}{4}\left({\hat\Omega}_{(-)}{}_{\mu}{}^{\hat a}{}_{\hat b}\,{\hat\Omega}_{(-)}{}_{y}{}^{\hat b}{}_{\hat a}-\frac{{\hat g}_{\mu y}{\hat\Omega}_{(-)}{}_{y}{}^{\hat a}{}_{\hat b}\,{\hat\Omega}_{(-)}{}_{y}{}^{\hat b}{}_{\hat a}}{{\hat g}_{yy}}\right)\,, \\[1mm]
\phi&=&\hat \phi-\frac{1}{2}\log\sqrt{-{\hat g}_{yy}}\, .
\end{eqnarray}
Finally, we define as in \cite{Elgood:2020xwu} the scalar combination
\begin{equation}\label{eq:scalarcombination}
k_{(1)}=k+\frac{\alpha'}{4}k^{-1}{\hat \Omega}_{(-)}{}_{y}{}^{\hat a}{}_{\hat b}{\hat \Omega}{(-)}_{y}{}^{\hat b}{}_{\hat a}\, .
\end{equation}

\bibliographystyle{JHEP}
\bibliography{references}

\providecommand{\href}[2]{#2}\begingroup\raggedright\begin{thebibliography}{10}

\bibitem{Dabholkar:1989jt}
A.~Dabholkar and J.A.~Harvey, \emph{{Nonrenormalization of the Superstring
  Tension}}, \href{https://doi.org/10.1103/PhysRevLett.63.478}{\emph{Phys. Rev.
  Lett.} {\bfseries 63} (1989) 478}.

\bibitem{Dabholkar:1990yf}
A.~Dabholkar, G.W.~Gibbons, J.A.~Harvey and F.~Ruiz~Ruiz, \emph{{Superstrings
  and Solitons}},
  \href{https://doi.org/10.1016/0550-3213(90)90157-9}{\emph{Nucl. Phys. B}
  {\bfseries 340} (1990) 33}.

\bibitem{Sen:1994eb}
A.~Sen, \emph{{Black hole solutions in heterotic string theory on a torus}},
  \href{https://doi.org/10.1016/0550-3213(95)00063-X}{\emph{Nucl. Phys. B}
  {\bfseries 440} (1995) 421}
  [\href{https://arxiv.org/abs/hep-th/9411187}{{\ttfamily hep-th/9411187}}].

\bibitem{Cvetic:1995uj}
M.~Cvetic and D.~Youm, \emph{{Dyonic BPS saturated black holes of heterotic
  string on a six torus}},
  \href{https://doi.org/10.1103/PhysRevD.53.R584}{\emph{Phys. Rev. D}
  {\bfseries 53} (1996) 584}
  [\href{https://arxiv.org/abs/hep-th/9507090}{{\ttfamily hep-th/9507090}}].

\bibitem{Dabholkar:1995nc}
A.~Dabholkar, J.P.~Gauntlett, J.A.~Harvey and D.~Waldram, \emph{{Strings as
  solitons and black holes as strings}},
  \href{https://doi.org/10.1016/0550-3213(96)00266-0}{\emph{Nucl. Phys. B}
  {\bfseries 474} (1996) 85}
  [\href{https://arxiv.org/abs/hep-th/9511053}{{\ttfamily hep-th/9511053}}].

\bibitem{Callan:1995hn}
C.G.~Callan, J.M.~Maldacena and A.W.~Peet, \emph{{Extremal black holes as
  fundamental strings}},
  \href{https://doi.org/10.1016/0550-3213(96)00315-X}{\emph{Nucl. Phys. B}
  {\bfseries 475} (1996) 645}
  [\href{https://arxiv.org/abs/hep-th/9510134}{{\ttfamily hep-th/9510134}}].

\bibitem{Sen:1995in}
A.~Sen, \emph{{Extremal black holes and elementary string states}},
  \href{https://doi.org/10.1142/S0217732395002234}{\emph{Mod. Phys. Lett. A}
  {\bfseries 10} (1995) 2081}
  [\href{https://arxiv.org/abs/hep-th/9504147}{{\ttfamily hep-th/9504147}}].

\bibitem{Dabholkar:2004yr}
A.~Dabholkar, \emph{{Exact counting of black hole microstates}},
  \href{https://doi.org/10.1103/PhysRevLett.94.241301}{\emph{Phys. Rev. Lett.}
  {\bfseries 94} (2005) 241301}
  [\href{https://arxiv.org/abs/hep-th/0409148}{{\ttfamily hep-th/0409148}}].

\bibitem{Dabholkar:2004dq}
A.~Dabholkar, R.~Kallosh and A.~Maloney, \emph{{A Stringy cloak for a classical
  singularity}},
  \href{https://doi.org/10.1088/1126-6708/2004/12/059}{\emph{JHEP} {\bfseries
  12} (2004) 059} [\href{https://arxiv.org/abs/hep-th/0410076}{{\ttfamily
  hep-th/0410076}}].

\bibitem{Cano:2018hut}
P.A.~Cano, P.F.~Ram\'\i{}rez and A.~Ruip\'erez, \emph{{The small black hole
  illusion}}, \href{https://doi.org/10.1007/JHEP03(2020)115}{\emph{JHEP}
  {\bfseries 03} (2020) 115}
  [\href{https://arxiv.org/abs/1808.10449}{{\ttfamily 1808.10449}}].

\bibitem{Ruiperez:2020qda}
A.~Ruip\'erez, \emph{{Higher-derivative corrections to small black rings}},
  \href{https://doi.org/10.1088/1361-6382/abff9b}{\emph{Class. Quant. Grav.}
  {\bfseries 38} (2021) 145011}
  [\href{https://arxiv.org/abs/2003.02269}{{\ttfamily 2003.02269}}].

\bibitem{Cano:2021dyy}
P.A.~Cano, A.~Murcia, P.F.~Ram\'\i{}rez and A.~Ruip\'erez, \emph{{On small
  black holes, KK monopoles and solitonic 5-branes}},
  \href{https://doi.org/10.1007/JHEP05(2021)272}{\emph{JHEP} {\bfseries 05}
  (2021) 272} [\href{https://arxiv.org/abs/2102.04476}{{\ttfamily
  2102.04476}}].

\bibitem{Susskind:1993ws}
L.~Susskind, \emph{{Some speculations about black hole entropy in string
  theory}},  \href{https://arxiv.org/abs/hep-th/9309145}{{\ttfamily
  hep-th/9309145}}.

\bibitem{Horowitz:1996nw}
G.T.~Horowitz and J.~Polchinski, \emph{{A Correspondence principle for black
  holes and strings}},
  \href{https://doi.org/10.1103/PhysRevD.55.6189}{\emph{Phys. Rev. D}
  {\bfseries 55} (1997) 6189}
  [\href{https://arxiv.org/abs/hep-th/9612146}{{\ttfamily hep-th/9612146}}].

\bibitem{Horowitz:1997jc}
G.T.~Horowitz and J.~Polchinski, \emph{{Selfgravitating fundamental strings}},
  \href{https://doi.org/10.1103/PhysRevD.57.2557}{\emph{Phys. Rev. D}
  {\bfseries 57} (1998) 2557}
  [\href{https://arxiv.org/abs/hep-th/9707170}{{\ttfamily hep-th/9707170}}].

\bibitem{Damour:1999aw}
T.~Damour and G.~Veneziano, \emph{{Selfgravitating fundamental strings and
  black holes}},
  \href{https://doi.org/10.1016/S0550-3213(99)00596-9}{\emph{Nucl. Phys. B}
  {\bfseries 568} (2000) 93}
  [\href{https://arxiv.org/abs/hep-th/9907030}{{\ttfamily hep-th/9907030}}].

\bibitem{Chen:2021emg}
Y.~Chen and J.~Maldacena, \emph{{String scale black holes at large D}},
  \href{https://doi.org/10.1007/JHEP01(2022)095}{\emph{JHEP} {\bfseries 01}
  (2022) 095} [\href{https://arxiv.org/abs/2106.02169}{{\ttfamily
  2106.02169}}].

\bibitem{Chen:2021dsw}
Y.~Chen, J.~Maldacena and E.~Witten, \emph{{On the black hole/string
  transition}}, \href{https://doi.org/10.1007/JHEP01(2023)103}{\emph{JHEP}
  {\bfseries 01} (2023) 103}
  [\href{https://arxiv.org/abs/2109.08563}{{\ttfamily 2109.08563}}].

\bibitem{Brustein:2021cza}
R.~Brustein and Y.~Zigdon, \emph{{Black hole entropy sourced by string winding
  condensate}}, \href{https://doi.org/10.1007/JHEP10(2021)219}{\emph{JHEP}
  {\bfseries 10} (2021) 219}
  [\href{https://arxiv.org/abs/2107.09001}{{\ttfamily 2107.09001}}].

\bibitem{Matsuo:2022kvx}
Y.~Matsuo, \emph{{Fluid model of a black hole-string transition}},
  \href{https://doi.org/10.1103/PhysRevD.107.126003}{\emph{Phys. Rev. D}
  {\bfseries 107} (2023) 126003}
  [\href{https://arxiv.org/abs/2205.15976}{{\ttfamily 2205.15976}}].

\bibitem{Balthazar:2022hno}
B.~Balthazar, J.~Chu and D.~Kutasov, \emph{{On Small Black Holes in String
  Theory}},  \href{https://arxiv.org/abs/2210.12033}{{\ttfamily 2210.12033}}.

\bibitem{Ceplak:2023afb}
N.~\v{C}eplak, R.~Emparan, A.~Puhm and M.~Toma\v{s}evi\'c, \emph{{The
  correspondence between rotating black holes and fundamental strings}},
  \href{https://arxiv.org/abs/2307.03573}{{\ttfamily 2307.03573}}.

\bibitem{Mathur:2018tib}
S.D.~Mathur and D.~Turton, \emph{{The fuzzball nature of two-charge black hole
  microstates}},
  \href{https://doi.org/10.1016/j.nuclphysb.2019.114684}{\emph{Nucl. Phys. B}
  {\bfseries 945} (2019) 114684}
  [\href{https://arxiv.org/abs/1811.09647}{{\ttfamily 1811.09647}}].

\bibitem{Callan:1988hs}
C.G.~Callan, Jr., R.C.~Myers and M.J.~Perry, \emph{{Black Holes in String
  Theory}}, \href{https://doi.org/10.1016/0550-3213(89)90172-7}{\emph{Nucl.
  Phys. B} {\bfseries 311} (1989) 673}.

\bibitem{Bergshoeff:1995cg}
E.~Bergshoeff, B.~Janssen and T.~Ortin, \emph{{Solution generating
  transformations and the string effective action}},
  \href{https://doi.org/10.1088/0264-9381/13/3/002}{\emph{Class. Quant. Grav.}
  {\bfseries 13} (1996) 321}
  [\href{https://arxiv.org/abs/hep-th/9506156}{{\ttfamily hep-th/9506156}}].

\bibitem{Kaloper:1997ux}
N.~Kaloper and K.A.~Meissner, \emph{{Duality beyond the first loop}},
  \href{https://doi.org/10.1103/PhysRevD.56.7940}{\emph{Phys. Rev. D}
  {\bfseries 56} (1997) 7940}
  [\href{https://arxiv.org/abs/hep-th/9705193}{{\ttfamily hep-th/9705193}}].

\bibitem{Bedoya:2014pma}
O.A.~Bedoya, D.~Marques and C.~Nunez, \emph{{Heterotic $\alpha$'-corrections in
  Double Field Theory}},
  \href{https://doi.org/10.1007/JHEP12(2014)074}{\emph{JHEP} {\bfseries 12}
  (2014) 074} [\href{https://arxiv.org/abs/1407.0365}{{\ttfamily 1407.0365}}].

\bibitem{Ortin:2020xdm}
T.~Ortin, \emph{{O(n, n) invariance and Wald entropy formula in the Heterotic
  Superstring effective action at first order in $\alpha'$}},
  \href{https://doi.org/10.1007/JHEP01(2021)187}{\emph{JHEP} {\bfseries 01}
  (2021) 187} [\href{https://arxiv.org/abs/2005.14618}{{\ttfamily
  2005.14618}}].

\bibitem{Eloy:2020dko}
C.~Eloy, O.~Hohm and H.~Samtleben, \emph{{Duality Invariance and Higher
  Derivatives}}, \href{https://doi.org/10.1103/PhysRevD.101.126018}{\emph{Phys.
  Rev. D} {\bfseries 101} (2020) 126018}
  [\href{https://arxiv.org/abs/2004.13140}{{\ttfamily 2004.13140}}].

\bibitem{Giveon:2009da}
A.~Giveon, D.~Gorbonos and M.~Stern, \emph{{Fundamental Strings and Higher
  Derivative Corrections to d-Dimensional Black Holes}},
  \href{https://doi.org/10.1007/JHEP02(2010)012}{\emph{JHEP} {\bfseries 02}
  (2010) 012} [\href{https://arxiv.org/abs/0909.5264}{{\ttfamily 0909.5264}}].

\bibitem{Cano:2018qev}
P.A.~Cano, P.~Meessen, T.~Ort\'\i{}n and P.F.~Ram\'\i{}rez,
  \emph{{$\alpha'$-corrected black holes in String Theory}},
  \href{https://doi.org/10.1007/JHEP05(2018)110}{\emph{JHEP} {\bfseries 05}
  (2018) 110} [\href{https://arxiv.org/abs/1803.01919}{{\ttfamily
  1803.01919}}].

\bibitem{Chimento:2018kop}
S.~Chimento, P.~Meessen, T.~Ortin, P.F.~Ramirez and A.~Ruiperez, \emph{{On a
  family of $\alpha'$-corrected solutions of the Heterotic Superstring
  effective action}},
  \href{https://doi.org/10.1007/JHEP07(2018)080}{\emph{JHEP} {\bfseries 07}
  (2018) 080} [\href{https://arxiv.org/abs/1803.04463}{{\ttfamily
  1803.04463}}].

\bibitem{Cano:2018brq}
P.A.~Cano, S.~Chimento, P.~Meessen, T.~Ort\'\i{}n, P.F.~Ram\'\i{}rez and
  A.~Ruip\'erez, \emph{{Beyond the near-horizon limit: Stringy corrections to
  Heterotic Black Holes}},
  \href{https://doi.org/10.1007/JHEP02(2019)192}{\emph{JHEP} {\bfseries 02}
  (2019) 192} [\href{https://arxiv.org/abs/1808.03651}{{\ttfamily
  1808.03651}}].

\bibitem{RuiperezVicente:2020qfw}
A.~Ruip\'erez~Vicente, \emph{{Black holes in string theory with
  higher-derivative corrections}}, Ph.D. thesis, U. Autonoma, Madrid (main),
  2020.

\bibitem{Elgood:2020nls}
Z.~Elgood, T.~Ort\'\i{}n and D.~Pere\~n\'\i{}guez, \emph{{The first law and
  Wald entropy formula of heterotic stringy black holes at first order in
  $\alpha'$}}, \href{https://doi.org/10.1007/JHEP05(2021)110}{\emph{JHEP}
  {\bfseries 05} (2021) 110}
  [\href{https://arxiv.org/abs/2012.14892}{{\ttfamily 2012.14892}}].

\bibitem{Elgood:2020xwu}
Z.~Elgood and T.~Ortin, \emph{{T duality and Wald entropy formula in the
  Heterotic Superstring effective action at first-order in
  \ensuremath{\alpha}'}},
  \href{https://doi.org/10.1007/JHEP10(2020)097}{\emph{JHEP} {\bfseries 10}
  (2020) 097} [\href{https://arxiv.org/abs/2005.11272}{{\ttfamily
  2005.11272}}].

\bibitem{Cano:2021nzo}
P.A.~Cano, T.~Ort\'\i{}n, A.~Ruip\'erez and M.~Zatti, \emph{{Non-supersymmetric
  black holes with \ensuremath{\alpha}' corrections}},
  \href{https://doi.org/10.1007/JHEP03(2022)103}{\emph{JHEP} {\bfseries 03}
  (2022) 103} [\href{https://arxiv.org/abs/2111.15579}{{\ttfamily
  2111.15579}}].

\bibitem{Ortin:2021win}
T.~Ort\'\i{}n, A.~Ruip\'erez and M.~Zatti, \emph{{Extremal stringy black holes
  in equilibrium at first order in $\alpha'$}},
  \href{https://arxiv.org/abs/2112.12764}{{\ttfamily 2112.12764}}.

\bibitem{Cano:2022tmn}
P.A.~Cano, T.~Ort\'\i{}n, A.~Ruip\'erez and M.~Zatti, \emph{{Non-extremal,
  \ensuremath{\alpha}'-corrected black holes in 5-dimensional heterotic
  superstring theory}},
  \href{https://doi.org/10.1007/JHEP12(2022)150}{\emph{JHEP} {\bfseries 12}
  (2022) 150} [\href{https://arxiv.org/abs/2210.01861}{{\ttfamily
  2210.01861}}].

\bibitem{Zatti:2023oiq}
M.~Zatti, \emph{{$\alpha'$ corrections to 4-dimensional non-extremal stringy
  black holes}},  \href{https://arxiv.org/abs/2308.12879}{{\ttfamily
  2308.12879}}.

\bibitem{Ortin:2015hya}
T.~Ortin, \emph{{Gravity and Strings}}, Cambridge Monographs on Mathematical
  Physics, Cambridge University Press, 2nd ed.~ed. (7, 2015),
  \href{https://doi.org/10.1017/CBO9781139019750}{10.1017/CBO9781139019750}.

\bibitem{Gross:1986mw}
D.J.~Gross and J.H.~Sloan, \emph{{The Quartic Effective Action for the
  Heterotic String}},
  \href{https://doi.org/10.1016/0550-3213(87)90465-2}{\emph{Nucl. Phys. B}
  {\bfseries 291} (1987) 41}.

\bibitem{Metsaev:1987zx}
R.R.~Metsaev and A.A.~Tseytlin, \emph{{Order alpha-prime (Two Loop) Equivalence
  of the String Equations of Motion and the Sigma Model Weyl Invariance
  Conditions: Dependence on the Dilaton and the Antisymmetric Tensor}},
  \href{https://doi.org/10.1016/0550-3213(87)90077-0}{\emph{Nucl. Phys. B}
  {\bfseries 293} (1987) 385}.

\bibitem{Bergshoeff:1989de}
E.A.~Bergshoeff and M.~de~Roo, \emph{{The Quartic Effective Action of the
  Heterotic String and Supersymmetry}},
  \href{https://doi.org/10.1016/0550-3213(89)90336-2}{\emph{Nucl. Phys. B}
  {\bfseries 328} (1989) 439}.

\bibitem{Chemissany:2007he}
W.A.~Chemissany, M.~de~Roo and S.~Panda, \emph{{alpha'-Corrections to Heterotic
  Superstring Effective Action Revisited}},
  \href{https://doi.org/10.1088/1126-6708/2007/08/037}{\emph{JHEP} {\bfseries
  08} (2007) 037} [\href{https://arxiv.org/abs/0706.3636}{{\ttfamily
  0706.3636}}].

\bibitem{Gibbons:1976ue}
G.W.~Gibbons and S.W.~Hawking, \emph{{Action Integrals and Partition Functions
  in Quantum Gravity}},
  \href{https://doi.org/10.1103/PhysRevD.15.2752}{\emph{Phys. Rev. D}
  {\bfseries 15} (1977) 2752}.

\bibitem{Page:1983mke}
D.N.~Page, \emph{{Classical Stability of Round and Squashed Seven Spheres in
  Eleven-dimensional Supergravity}},
  \href{https://doi.org/10.1103/PhysRevD.28.2976}{\emph{Phys. Rev. D}
  {\bfseries 28} (1983) 2976}.

\bibitem{Marolf:2000cb}
D.~Marolf, \emph{{Chern-Simons terms and the three notions of charge}},  in
  \emph{{International Conference on Quantization, Gauge Theory, and Strings:
  Conference Dedicated to the Memory of Professor Efim Fradkin}}, pp.~312--320,
  6, 2000 [\href{https://arxiv.org/abs/hep-th/0006117}{{\ttfamily
  hep-th/0006117}}].

\bibitem{Maldacena:1996ky}
J.M.~Maldacena, \emph{{Black holes in string theory}}, Ph.D. thesis, Princeton
  U., 1996.
\newblock \href{https://arxiv.org/abs/hep-th/9607235}{{\ttfamily
  hep-th/9607235}}.

\bibitem{Wald:1993nt}
R.M.~Wald, \emph{{Black hole entropy is the Noether charge}},
  \href{https://doi.org/10.1103/PhysRevD.48.R3427}{\emph{Phys. Rev. D}
  {\bfseries 48} (1993) R3427}
  [\href{https://arxiv.org/abs/gr-qc/9307038}{{\ttfamily gr-qc/9307038}}].

\bibitem{Iyer:1994ys}
V.~Iyer and R.M.~Wald, \emph{{Some properties of Noether charge and a proposal
  for dynamical black hole entropy}},
  \href{https://doi.org/10.1103/PhysRevD.50.846}{\emph{Phys. Rev. D} {\bfseries
  50} (1994) 846} [\href{https://arxiv.org/abs/gr-qc/9403028}{{\ttfamily
  gr-qc/9403028}}].

\bibitem{Cassani:2023vsa}
D.~Cassani, A.~Ruip\'erez and E.~Turetta, \emph{{Boundary terms and conserved
  charges in higher-derivative gauged supergravity}},
  \href{https://doi.org/10.1007/JHEP06(2023)203}{\emph{JHEP} {\bfseries 06}
  (2023) 203} [\href{https://arxiv.org/abs/2304.06101}{{\ttfamily
  2304.06101}}].

\bibitem{Sahoo:2006vz}
B.~Sahoo and A.~Sen, \emph{{BTZ black hole with Chern-Simons and higher
  derivative terms}},
  \href{https://doi.org/10.1088/1126-6708/2006/07/008}{\emph{JHEP} {\bfseries
  07} (2006) 008} [\href{https://arxiv.org/abs/hep-th/0601228}{{\ttfamily
  hep-th/0601228}}].

\bibitem{Sahoo:2006pm}
B.~Sahoo and A.~Sen, \emph{{alpha-prime - corrections to extremal dyonic black
  holes in heterotic string theory}},
  \href{https://doi.org/10.1088/1126-6708/2007/01/010}{\emph{JHEP} {\bfseries
  01} (2007) 010} [\href{https://arxiv.org/abs/hep-th/0608182}{{\ttfamily
  hep-th/0608182}}].

\bibitem{Faedo:2019xii}
F.~Faedo and P.F.~Ramirez, \emph{{Exact charges from heterotic black holes}},
  \href{https://doi.org/10.1007/JHEP10(2019)033}{\emph{JHEP} {\bfseries 10}
  (2019) 033} [\href{https://arxiv.org/abs/1906.12287}{{\ttfamily
  1906.12287}}].

\bibitem{Ma:2022nwq}
L.~Ma, Y.~Pang and H.~Lu, \emph{{Improved Wald formalism and first law of
  dyonic black strings with mixed Chern-Simons terms}},
  \href{https://doi.org/10.1007/JHEP10(2022)142}{\emph{JHEP} {\bfseries 10}
  (2022) 142} [\href{https://arxiv.org/abs/2202.08290}{{\ttfamily
  2202.08290}}].

\bibitem{Tachikawa:2006sz}
Y.~Tachikawa, \emph{{Black hole entropy in the presence of Chern-Simons
  terms}}, \href{https://doi.org/10.1088/0264-9381/24/3/014}{\emph{Class.
  Quant. Grav.} {\bfseries 24} (2007) 737}
  [\href{https://arxiv.org/abs/hep-th/0611141}{{\ttfamily hep-th/0611141}}].

\bibitem{Reall:2019sah}
H.S.~Reall and J.E.~Santos, \emph{{Higher derivative corrections to Kerr black
  hole thermodynamics}},
  \href{https://doi.org/10.1007/JHEP04(2019)021}{\emph{JHEP} {\bfseries 04}
  (2019) 021} [\href{https://arxiv.org/abs/1901.11535}{{\ttfamily
  1901.11535}}].

\bibitem{Bobev:2022bjm}
N.~Bobev, V.~Dimitrov, V.~Reys and A.~Vekemans, \emph{{Higher derivative
  corrections and AdS5 black holes}},
  \href{https://doi.org/10.1103/PhysRevD.106.L121903}{\emph{Phys. Rev. D}
  {\bfseries 106} (2022) L121903}
  [\href{https://arxiv.org/abs/2207.10671}{{\ttfamily 2207.10671}}].

\bibitem{Cassani:2022lrk}
D.~Cassani, A.~Ruip\'erez and E.~Turetta, \emph{{Corrections to AdS$_{5}$ black
  hole thermodynamics from higher-derivative supergravity}},
  \href{https://doi.org/10.1007/JHEP11(2022)059}{\emph{JHEP} {\bfseries 11}
  (2022) 059} [\href{https://arxiv.org/abs/2208.01007}{{\ttfamily
  2208.01007}}].

\bibitem{Baron:2017dvb}
W.H.~Baron, J.J.~Fernandez-Melgarejo, D.~Marques and C.~Nunez, \emph{{The Odd
  story of \ensuremath{\alpha}'-corrections}},
  \href{https://doi.org/10.1007/JHEP04(2017)078}{\emph{JHEP} {\bfseries 04}
  (2017) 078} [\href{https://arxiv.org/abs/1702.05489}{{\ttfamily
  1702.05489}}].

\bibitem{Liu:2023fqq}
J.T.~Liu and R.J.~Saskowski, \emph{{Consistent truncations in higher derivative
  supergravity}},  \href{https://arxiv.org/abs/2307.12420}{{\ttfamily
  2307.12420}}.

\bibitem{Tseytlin:1988tv}
A.A.~Tseytlin, \emph{{Mobius Infinity Subtraction and Effective Action in
  $\sigma$ Model Approach to Closed String Theory}},
  \href{https://doi.org/10.1016/0370-2693(88)90421-2}{\emph{Phys. Lett. B}
  {\bfseries 208} (1988) 221}.

\bibitem{Green:1984sg}
M.B.~Green and J.H.~Schwarz, \emph{{Anomaly Cancellation in Supersymmetric D=10
  Gauge Theory and Superstring Theory}},
  \href{https://doi.org/10.1016/0370-2693(84)91565-X}{\emph{Phys. Lett. B}
  {\bfseries 149} (1984) 117}.

\end{thebibliography}\endgroup
\end{document}